%% file: cccoherentpi.tex
\documentclass[twocolumn,showpacs,preprintnumbers,amsmath,amssymb,superscriptaddress]{revtex4}

\input{preamble}

\begin{document}

\title{\bf Search for Charged Current Coherent Pion Production on Carbon in a Few-GeV Neutrino Beam}
\input{authors}

\date{\today}

\input{abstract}


\maketitle

\input{introduction}

\input{beam}

\input{neutrinointeractionmc}

\input{detector}

\input{data}

\input{analysis}

\input{results}

\input{conclusions}

\input{acknowledgments}

\end{document}

%% file: preamble.tex
\usepackage{graphicx}
\usepackage{dcolumn}
\usepackage{bm}
\usepackage[figuresright]{rotating}
\usepackage{fancyhdr}
\usepackage{enumerate}
\usepackage{indentfirst}
\usepackage{xspace}
\usepackage{color}

%% file: authors.tex
\newcommand{\BARCELONA}{\affiliation{Institut de Fisica d'Altes Energies, Universitat Autonoma de Barcelona, E-08193 Bellaterra (Barcelona), Spain}}
\newcommand{\COLORADO}{\affiliation{Department of Physics, University of Colorado, Boulder, Colorado 80309, USA}}
\newcommand{\COLUMBIA}{\affiliation{Department of Physics, Columbia University, New York, NY 10027, USA}}
\newcommand{\FNAL}{\affiliation{Fermi National Accelerator Laboratory; Batavia, IL 60150, USA}}
\newcommand{\KEK}{\affiliation{High Energy Accelerator Research Organization (KEK), Tsukuba, Ibaraki 305-0801, Japan}}
\newcommand{\IMPERIAL}{\affiliation{Department of Physics, Imperial College London, London SW7 2AZ, UK}}
\newcommand{\INDIANA}{\affiliation{Department of Physics, Indiana University, Bloomington, IN 47405, USA}}
\newcommand{\ICRRKAM}{\affiliation{Kamioka Observatory, Institute for Cosmic Ray Research, University of Tokyo, Gifu 506-1205, Japan}}
\newcommand{\ICRRKAS}{\affiliation{Research Center for Cosmic Neutrinos, Institute for Cosmic Ray Research, University of Tokyo, Kashiwa, Chiba 277-8582, Japan}}
\newcommand{\KYOTO}{\affiliation{Department of Physics, Kyoto University, Kyoto 606-8502, Japan}}
\newcommand{\LANL}{\affiliation{Los Alamos National Laboratory; Los Alamos, NM 87545, USA}}
\newcommand{\LSU}{\affiliation{Department of Physics and Astronomy, Louisiana State University, Baton Rouge, LA 70803, USA}}
\newcommand{\MIT}{\affiliation{Department of Physics, Massachusetts Institute of Technology, Cambridge, MA 02139, USA}}
\newcommand{\PURDUE}{\affiliation{Department of Chemistry and Physics, Purdue University Calumet, Hammond, IN 46323, USA}}
\newcommand{\ROMA}{\affiliation{Universita di Roma La Sapienza, Dipartmento di Fisica and INFN, 1-000185 Rome, Italy}}
\newcommand{\STMARYS}{\affiliation{Physics Department, Saint Mary's University of Minnesota, Winona, MN 55987, USA}}
\newcommand{\TOKYOT}{\affiliation{Department of Physics, Tokyo Institute of Technology, Tokyo 152-8551, Japan}}
\newcommand{\VALENCIA}{\affiliation{Instituto de Fisica Corpuscular, Universidad de Valencia and CSIC, E-46071 Valencia, Spain}}

\BARCELONA
\COLORADO
\COLUMBIA
\FNAL
\KEK
\IMPERIAL
\INDIANA
\ICRRKAM
\ICRRKAS
\KYOTO
\LANL
\LSU
\MIT
\PURDUE
\ROMA
\STMARYS
\TOKYOT
\VALENCIA

\author{K.~Hiraide}\KYOTO
\author{J.~L.~Alcaraz-Aunion}\BARCELONA
\author{S.~J.~Brice}\FNAL
\author{L.~Bugel}\MIT
\author{J.~Catala-Perez}\VALENCIA
\author{G.~Cheng}\COLUMBIA
\author{J.~M.~Conrad}\MIT
\author{Z.~Djurcic}\COLUMBIA
\author{U.~Dore}\ROMA
\author{D.~A.~Finley}\FNAL
\author{A.~J.~Franke}\COLUMBIA
\author{C.~Giganti\footnote{Present address: DSM/Irfu/SPP, CEA Saclay, F-91191 Gif-sur-Yvette, France}}\ROMA
\author{J.~J.~Gomez-Cadenas}\VALENCIA
\author{P.~Guzowski}\IMPERIAL
\author{A.~Hanson}\INDIANA
\author{Y.~Hayato}\ICRRKAM
\author{G.~Jover-Manas}\BARCELONA
\author{G.~Karagiorgi}\MIT
\author{T.~Katori}\INDIANA
\author{Y.~K.~Kobayashi}\TOKYOT
\author{T.~Kobilarcik}\FNAL
\author{H.~Kubo}\KYOTO
\author{Y.~Kurimoto}\KYOTO
\author{W.~C.~Louis}\LANL
\author{P.~F.~Loverre}\ROMA
\author{L.~Ludovici}\ROMA
\author{K.~B.~M.~Mahn}\COLUMBIA
\author{C.~Mariani\footnote{Present address: Department of Physics, Columbia University, New York, NY 10027, USA}}\ROMA
\author{S.~Masuike}\TOKYOT
\author{K.~Matsuoka}\KYOTO
\author{W.~Metcalf}\LSU
\author{G.~Mills}\LANL
\author{G.~Mitsuka}\ICRRKAS
\author{Y.~Miyachi}\TOKYOT
\author{S.~Mizugashira}\TOKYOT
\author{C.~D.~Moore}\FNAL
\author{Y.~Nakajima}\KYOTO
\author{T.~Nakaya}\KYOTO
\author{R.~Napora}\PURDUE
\author{P.~Nienaber}\STMARYS
\author{V.~Nguyen}\MIT
\author{D.~Orme}\KYOTO
\author{M.~Otani}\KYOTO
\author{A.~D.~Russell}\FNAL
\author{F.~Sanchez}\BARCELONA
\author{M.~H.~Shaevitz}\COLUMBIA
\author{T.-A.~Shibata}\TOKYOT
\author{M.~Sorel}\VALENCIA
\author{R.~J.~Stefanski}\FNAL
\author{H.~Takei}\TOKYOT
\author{H.-K.~Tanaka}\COLUMBIA
\author{M.~Tanaka}\KEK
\author{R.~Tayloe}\INDIANA
\author{I.~J.~Taylor}\IMPERIAL
\author{R.~J.~Tesarek}\FNAL
\author{Y.~Uchida}\IMPERIAL
\author{R.~Van~de~Water}\LANL
\author{J.~J.~Walding}\IMPERIAL
\author{M.~O.~Wascko}\IMPERIAL
\author{H.~White}\FNAL
\author{M.~J.~Wilking}\COLORADO
\author{M.~Yokoyama}\KYOTO
\author{G.~P.~Zeller}\LANL
\author{E.~D.~Zimmerman}\COLORADO

\collaboration{The SciBooNE Collaboration}\noaffiliation

%% file: abstract.tex
\begin{abstract}

The SciBooNE Collaboration has performed a search for charged current
coherent pion production from muon neutrinos scattering on carbon,
${\nu_\mu}^{12}{\rm C} \to \mu^{- 12}{\rm C} \pi^+$, with
two distinct data samples.  No evidence for coherent pion production
is observed.  We set 90\% confidence level upper limits on the cross
section ratio of charged current coherent pion production to the total
charged current cross section at $0.67\times 10^{-2}$ at mean neutrino
energy 1.1~GeV and $1.36\times 10^{-2}$ at mean neutrino energy
2.2~GeV.

\end{abstract}

%% file: introduction.tex
\section{Introduction}
\label{sec:introduction}

Although they have been studied for decades, neutrino-nucleus cross
sections between 100~MeV and 10~GeV energy are still known with very
poor accuracy.  The demand for precise cross section measurements in
this energy regime is driven by the needs of the next generation of
neutrino oscillation experiments in their pursuit of sub-leading
flavor oscillation and charge-parity violation
\cite{Itow:2002rk,Harris:2004iq}.  Several interaction channels
contribute to the total neutrino-nucleus cross section in the neutrino
energy range of a few GeV.  Interactions producing single pions
(charged or neutral) account for a large cross section fraction, which
must be understood because they form significant backgrounds for
neutrino oscillation searches.

It has been known for years that neutrinos can produce pions by
interacting {\it coherently} with the nucleons forming the target
nucleus. The cross section for this process is expected to be smaller
than incoherent pion production, the latter being dominated by
neutrino-induced baryonic resonance excitation off a single nucleon
bound in a nucleus. Moreover, coherent pion production is
comparatively poorly understood, although it is characterized by a
distinct signature consisting of a nucleus left in the ground state
(no nuclear breakup occurs) and a forward scattered pion. Both
charged current and neutral current coherent modes are
possible, $\nu_{\mu}A\to\mu^-A\pi^+$ and
$\nu_{\mu}A\to\nu_{\mu}A\pi^0$, where $A$ is a nucleus.

Several theoretical models describing coherent pion production have
been proposed, using different formalisms to describe the relevant
physics. A first class of models is built on the basis of Adler's PCAC
theorem~\cite{Adler:1964yx}, relating the neutrino-nucleus cross
section to that of a pion interacting with a nucleus at $Q^2=0$; the
extrapolation to $Q^2\neq 0$ is performed via a propagator term
\cite{Gershtein:1980vd,Rein:1982pf,Belkov:1986hn,Paschos:2005km,Rein:2006di}.
A second commonly-used formalism is based on the description of the coherent
production of $\Delta$ resonances on nuclei by using a modified
$\Delta$-propagator and a distorted wave-function for the pion
\cite{Singh:2006bm,AlvarezRuso:2007tt,AlvarezRuso:2007it,Amaro:2008hd}. While the
relationship between neutral current and charged current modes, and that between neutrino and
antineutrino coherent pion production cross sections, are relatively
well known, order-of-magnitude variations on absolute coherent pion
production cross sections are expected within these models. In
addition, the cross section dependence on neutrino energy and on
target material is also uncertain. It is therefore imperative that
more experimental input on coherent pion production in
neutrino-nucleus interactions is gathered in the near future.

Coherent pion production in neutrino-nucleus interactions has already
been the subject of several experimental campaigns. The neutrino
energy range between 1 and 100~GeV has been investigated, including
both the charged current and neutral current modes, and using both neutrino and antineutrino
probes. A result that has drawn much attention in the neutrino physics
community has been the recent non-observation of charged current coherent pion
production by the K2K experiment with a 1.3~GeV wide-band neutrino
beam \cite{Hasegawa:2005td}. This is motivated by the fact that the
K2K Collaboration has quoted an upper limit for the ratio of the charged current
coherent pion production cross section to the charged current inclusive
cross section that is well below the prediction of the original
Rein-Sehgal model \cite{Rein:1982pf} that has been adopted in the past
to describe coherent pion production processes. In addition, even
within more recently proposed models, it is often difficult to
reconcile this new and accurate null result at low energies with
previous measurements. On the one hand, evidence for neutral current coherent pion
production in a neutrino energy range that is similar to K2K has been
unambiguously reported from the Aachen-Padova \cite{Faissner:1983ng}
and Gargamelle experimental data \cite{Isiksal:1984vh} first, and more
recently also by the MiniBooNE Collaboration
\cite{AguilarArevalo:2008xs}. On the other hand, while no measurements
of charged current coherent pion production other than the K2K one exist in a
similar neutrino energy range, there exist charged current coherent pion production
positive results at higher energies (7-100~GeV neutrino energy) from
the SKAT \cite{Grabosch:1985mt}, CHARM \cite{Vilain:1993sf}, BEBC
\cite{Marage:1986cy,Allport:1988cq}, and FNAL E632
\cite{Willocq:1992fv} experiments.

In this paper, we discuss the first measurement of charged current coherent pion
production by neutrinos in the SciBooNE experiment
\cite{AguilarArevalo:2006se}. This is a particularly interesting test
of the K2K null result, probing a similar neutrino energy range and 
same target material. Also,
compared to K2K, SciBooNE's result presented here is based on a
higher-statistics data sample and uses an improved analysis, as will
be described below.

This paper is organized as follows.  Section \ref{sec:beam} describes
the neutrino beam-line and the neutrino flux simulation.  The
simulation of neutrino interactions with nuclei are described in
Section \ref{sec:neutrinointeractionmc}.  The detector configuration
and simulation are described in Section \ref{sec:detector}.  A summary
of the data set and experimental performance is given in Section
\ref{sec:data}.  The data analysis, including the event selection and
Monte Carlo (MC) tuning, is described in detail in Section
\ref{sec:analysis}.  The results of the analysis and discussion are
presented along with a summary of systematic uncertainties in Section
\ref{sec:results}, and the final conclusions are given in Section
\ref{sec:conclusions}.

%% file: beam.tex
\section{Neutrino Beam}
\label{sec:beam}

The SciBooNE detector has been exposed to the Booster Neutrino Beam
(BNB) located at Fermilab. The BNB is a high-intensity, conventional
neutrino beam which has been serving the MiniBooNE experiment since
2002.

\subsection{Beam-line Description}
\label{sec:beamline}

The primary beam uses protons accelerated to 8 GeV kinetic energy by
the Fermilab Booster. Selected batches containing approximately
4-5$\times 10^{12}$ protons are extracted
and bent toward the BNB target hall via dipole magnets. Each spill is
composed of 81 bunches of protons, approximately 6~ns wide each and 19~ns
apart, for a total spill duration of $1.6\ \mu$s.

Beam proton trajectories and positions are monitored on a
pulse-by-pulse basis.
The typical beam alignment and divergence measured by the beam
position monitors located near the target are within 1 mm and 1 mrad
of the nominal target center and axis direction, respectively; the
typical beam focusing on target measured by beam profile monitors is
of the order of 1-2 mm (RMS) in both the horizontal and vertical
directions. These parameters are well within the experiment
requirements. The number of protons delivered to the BNB target is
measured for each proton batch using two toroids located near the
target along the beam-line. The toroid calibration, performed on a
pulse-by-pulse basis, provides a measurement of the number of
protons to BNB with a 2\% accuracy.

Primary protons from the 8 GeV beamline strike a thick beryllium
target located in the BNB target hall. Hadronic interactions of the
protons with the target material produce a beam of secondary mesons
(pions and kaons). The target is made of seven cylindrical slugs
for a total target length of 71.1~cm, or about 1.7 inelastic
interaction lengths.

The beryllium target is surrounded by a magnetic focusing horn,
bending and sign-selecting the secondary particles that emerge from
the interactions in the target along the direction pointing to the
SciBooNE detector. The focusing is produced by the toroidal magnetic
field present in the air volume between the horn's two coaxial
conductors made of aluminum alloy.
The horn current pulse is approximately a half-sinusoid of amplitude
174~kA, 143 $\mu$s long, synchronized to each beam spill.
The polarity of the horn current flow can be (and has been) switched,
in order to focus negatively-charged mesons, and therefore produce an
antineutrino instead of a neutrino beam.

The beam of focused, secondary mesons emerging from the target/horn
region is further collimated via passive shielding, and allowed to
decay into neutrinos in a cylindrical decay region filled with air at
atmospheric pressure, 50 m long and 90 cm in radius. A beam absorber
located at the end of the decay region stops the hadronic and muonic
component of the beam, and only a pure neutrino beam pointing toward
the detector remains, mostly from $\pi^+\to\mu^+ \nu_{\mu}$ decays.

\subsection{Neutrino Flux Prediction}
\label{sec:fluxprediction}

Neutrino flux predictions at the SciBooNE detector location are
obtained via a GEANT4-based \cite{Agostinelli:2002hh} beam Monte Carlo
simulation. The same simulation code developed by the MiniBooNE
Collaboration is used \cite{AguilarArevalo:2008yp}.

In the simulation code, a realistic description of the geometry and
materials present in the BNB target hall and decay region is
used. Primary protons are generated according to the expected beam
optics properties upstream of the target. The interactions of primary
protons with the beryllium target are simulated according to
state-of-the-art hadron interaction data. Of particular importance for
this analysis is $\pi^+$ production in proton-beryllium interactions,
which uses experimental input from the HARP \cite{:2007gt} and BNL
E910 \cite{:2007nb} experiments.  Production of secondary protons,
neutrons, charged pions, and charged and neutral kaons is taken into
account, and elastic and quasi-elastic scattering of protons in the
target are also simulated. Particles emanating from the primary
proton-beryllium interaction in the target are then propagated within
the GEANT4 framework, which accounts for all relevant physics
processes. Hadronic re-interactions of pions and nucleons with
beryllium and aluminum materials are particularly important and are
described by custom models, while other hadronic processes and all
electromagnetic processes (energy loss, multiple scattering, effect of
horn magnetic field, etc.) are described according to default GEANT4
physics lists. A second, FORTRAN-based Monte Carlo code uses the
output of the GEANT4 program as input, and is responsible for
generating the neutrino kinematics distributions from meson and muon
decays, and for obtaining the final neutrino fluxes extrapolated to
the SciBooNE detector with negligible beam Monte Carlo statistical
errors. Current best knowledge of neutrino-producing meson and muon
decay branching fractions, and decay form factors in three-body
semi-leptonic decays, are used. Polarization effects in muon decays are
also accounted for.

Once produced by the simulation, neutrinos are extrapolated along
straight lines toward the SciBooNE detector. All neutrinos whose ray
traces cross any part of the detector volume are considered for
SciBooNE flux predictions. Based on accurate survey data, the distance
between the center of the beryllium target and the center of the
SciBar detector is taken to be 99.9 m, with the SciBooNE detector
located on beam axis within a tolerance of a few cm.
Each simulated neutrino interaction is linked to its detailed beam
information and history, which includes neutrino flavor, energy,
parent type and kinematics, and ray trace entry and exit points within
the detector volume; the ray trace information is used to determine
the incoming neutrino's direction and interaction location. Proper
weights for each beam neutrino event are computed, using this
beam neutrino information, as well as information from the
interaction and detector simulation: neutrino interaction probability,
and detailed SciBooNE detector geometry and specifications.
\begin{figure}[ht]
\begin{center}
  \includegraphics[width=\columnwidth]{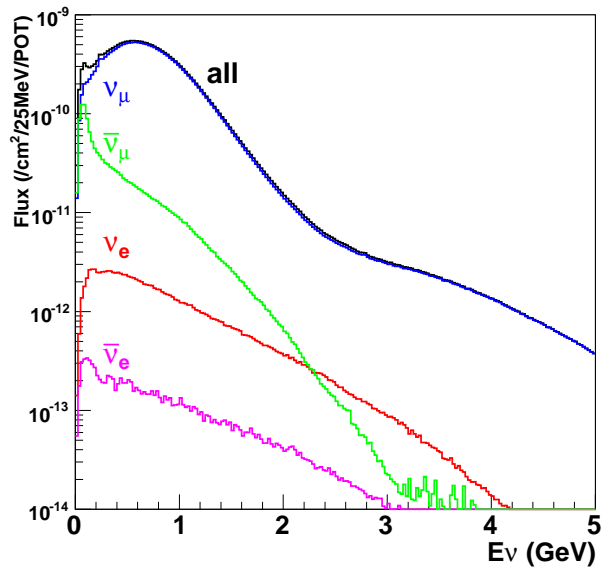}
\caption{Neutrino flux prediction at the SciBooNE detector as a
  function of neutrino energy $E_{\nu}$, normalized per unit area,
  proton on target (POT) and neutrino energy bin width.
  The spectrum is averaged within 2.12~m from the beam center.
  The total flux and contributions from individual neutrino flavors
  are shown.
  }
\label{fig:flux}
\end{center}
\end{figure}

The neutrino flux prediction at the SciBooNE detector location and as
a function of neutrino energy is shown in Fig.~\ref{fig:flux}. A total
neutrino flux per proton on target of $2.2\times 10^{-8}$~cm$^{-2}$ is
expected at the SciBooNE detector location and in neutrino running
mode (positive horn polarity), with a mean neutrino energy of 0.7
GeV. The flux is dominated by muon neutrinos (93\% of total), with
small contributions from muon antineutrinos (6.4\%), and electron
neutrinos and antineutrinos (0.6\% in total). For the neutrino flux
predictions used in this analysis, no information from BNB (SciBooNE
or MiniBooNE) neutrino data is used as experimental input.

%% file: neutrinointeractionmc.tex
\section{Neutrino Interaction Simulation}
\label{sec:neutrinointeractionmc}

The neutrino interactions with nuclear targets are simulated with the
NEUT program library~\cite{Hayato:2002sd,Mitsuka:2008zz} which is used
in the Kamiokande, Super-Kamiokande, K2K, and T2K experiments.  NEUT
handles protons, oxygen, carbon, and iron as nuclear targets in the
energy range from 100~MeV to 100~TeV.  In NEUT, the following neutrino
interactions in both neutral and charged currents are simulated:
quasi-elastic scattering ($\nu N \rightarrow \ell N'$), single meson
production ($\nu N \rightarrow \ell N'm$), single gamma production
($\nu N \rightarrow \ell N' \gamma$), coherent $\pi$
production ($\nu^{12}{\rm C (or } ^{56}{\rm Fe})
\rightarrow \ell\pi\ ^{12}{\rm C (or } ^{56}{\rm Fe})$), and deep
inelastic scattering ($\nu N \rightarrow \ell N'hadrons$), where
$N$ and $N'$ are the nucleons (proton or neutron), $\ell$ is the
lepton, and $m$ is the meson. Following the primary neutrino
interactions in nuclei, re-interactions of the mesons and hadrons with the
nuclear medium are also simulated.

\subsection{Coherent $\pi$ production}
\begin{figure}[bhtp]
  \includegraphics[height=8cm,angle=270]{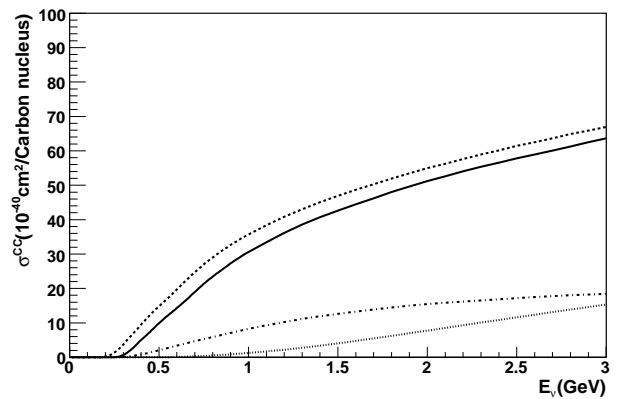}
  \caption{Cross section for ${\nu_{\mu}}^{12} $C$ \to \mu^-\pi^{+ 12}$C
  interaction. The solid line represents the Rein and
  Sehgal model with lepton mass effects~\cite{Rein:2006di}(the default
  model of the signal for this analysis), the dashed line represents
  the Rein and Sehgal model without lepton mass effects~\cite{Rein:1982pf},
  the dotted line represents the model of  Kartavtsev {\it et al}.~\cite{Paschos:2005km},
  and the dashed-dotted line represents the model of Alvarez-Ruso
  {\it et al}.~\cite{AlvarezRuso:2007it}.
  The model of Singh {\it et al}.~\cite{Singh:2006bm} gives a cross 
  section similar to the model of Alvarez-Ruso {\it et al}.
  }
  \label{fig:coherent}
\end{figure}

The signal for this analysis, coherent pion production, is a neutrino
interaction with a nucleus which remains intact, releasing one pion
with the same charge as the incoming weak current. Because of the
small momentum transfer to the target nucleus the outgoing lepton and
pion tend to go in the forward direction (in the lab frame). The
formalism developed by Rein and Sehgal~\cite{Rein:1982pf} is used to
simulate the interactions, including the recent correction of lepton
mass effects~\cite{Rein:2006di}.
The axial vector mass, $M_A$, is set to 1.0~GeV/$c^2$.
The nuclear radius parameter $R_0$ is set to 1.0~fm.
For the total and inelastic pion-nucleon cross sections used in the formalism,
the fitted results given in Rein and Sehgal's paper are employed.
The total cross section on $^{12}$C is shown in Fig.~\ref{fig:coherent},
with comparisons of other models
discussed in the introduction.  The Rein and Sehgal model predicts
charged current coherent pion production to be approximately 1\% of the total
neutrino interactions in SciBooNE.

\subsection{Quasi-elastic scattering}
The dominant interaction in the SciBooNE neutrino energy range is
quasi-elastic scattering, which is  implemented using the model of
Llewellyn-Smith~\cite{Llewellyn Smith:1971zm}. For scattering off
nucleons in the nucleus, we use the relativistic Fermi gas model of
Smith and Moniz~\cite{Smith:1972xh}.  The nucleons are treated as
quasi-free particles and the Fermi motion of nucleons along with the
Pauli exclusion principle is taken into account.  The momentum
distribution of the target nucleon is assumed to be flat up to a fixed
Fermi surface momentum of 217~MeV/$c$ for carbon and 250~MeV/$c$ for
iron. The same Fermi momentum distribution is also used for all of the
other nuclear interactions. The nuclear potential is set to 27~MeV for
carbon and 32~MeV for iron.  
Both vector and axial-vector form factor are assumed to be dipole. 
The vector mass in quasi-elastic scattering is set to be 0.84~GeV/$c^2$.
The axial vector mass, $M_A$, is set to be 1.21~GeV/$c^2$ as suggested
by recent results~\cite{Gran:2006jn,:2007ru}.

\subsection{Single meson production via baryon resonances}
The second most probable interaction in SciBooNE is the resonant
single meson production of $\pi$, $K$, and $\eta$ described by the
model of Rein and Sehgal~\cite{Rein:1980wg}.
The model assumes an intermediate baryon resonance, $N^*$,
in the reaction of $\nu N \rightarrow \ell N^*, N^* \rightarrow N'm$. The
differential cross section of single meson production depends on the
amplitude for the production of a given resonance and the probability
of the baryon resonance decay to the meson.  All intermediate baryon
resonances with mass less than 2~GeV/$c^2$ are included.  Those baryon
resonances with mass greater than 2~GeV/$c^2$ are simulated as deep
inelastic scattering.  Lepton mass effects from the non-conservation of
lepton current and the pion-pole term in the hadronic axial vector
current are included in the
simulation~\cite{Berger:2007rq,Kuzmin:2003ji}.

To determine the angular distribution of a pion in the final state,
Rein's method~\cite{Rein:1987cb} is used for the $P_{33}(1232)$ resonance. For
other resonances, the directional distribution of the generated pion
is set to be isotropic in the resonance rest frame.
The angular distribution of $\pi^+$ has been measured for
$\nu_\mu p \to \mu^- p \pi^+$~\cite{Kitagaki:1986ct} and the results
agree well with NEUT's prediction.
Pauli blocking is accounted for in the decay of the
baryon resonance by requiring the momentum of the nucleon to be larger
than the Fermi surface momentum. Pion-less $\Delta$ decay is also
taken into account, where 20\% of the events do not have a pion
and only the lepton and nucleon are generated~\cite{Singh:1998ha}.
The axial vector mass, $M_A$, is set to be 1.21~GeV/$c^2$.


\subsection{Deep inelastic scattering}
The cross section for deep inelastic scattering (DIS) is calculated using
the GRV98 parton distribution functions~\cite{Gluck:1998xa}.
Additionally, we have included the corrections in the small $Q^2$
region developed by Bodek and Yang~\cite{Bodek:2003wd}. In the
calculation, the hadronic invariant mass, $W$, is required to be
larger than 1.3~GeV/$c^2$. Also, the multiplicity of pions is
restricted to be larger than or equal to two for $1.3<W<2.0~{\rm
  GeV}/c^2$, because single pion production is already included in the
simulation, as described above.  The multi-hadron final states are
simulated with two models: a custom-made
program~\cite{Nakahata:1986zp} for the event with $W$ between 1.3 and
2.0~GeV/$c^2$ and PYTHIA/JETSET~\cite{Sjostrand:1993yb} for the events
with $W$ larger than 2~GeV/$c^2$.

\subsection{Intra-nuclear interactions}
\label{sec:intra_nuclear}
The intra-nuclear interactions of mesons and nucleons produced in
neutrino interactions in the nuclei are simulated. These interactions
are treated using a cascade model, and each of the particles is traced
until it escapes from the nucleus.

Among all the interactions of mesons and nucleons, the interactions of
pions are most important to this analysis. The inelastic scattering,
charge exchange and absorption of pions in the nuclei are
simulated. The interaction cross sections of pions in the nuclei
are calculated using the model by Salcedo {\it et al}.~\cite{Salcedo:1987md},
which agrees well with past experimental data~\cite{Ashery:1981tq}.
If inelastic scattering or charge exchange occurs, the
direction and momentum of pions are determined by using results from 
a phase shift analysis of pion-nucleus scattering experiments~\cite{Rowe:1978fb}.
When calculating the pion scattering amplitude, Pauli
blocking is taken into account by requiring the nucleon momentum after
the interaction to be larger than the Fermi surface momentum at the
interaction point.

Re-interactions of the recoil protons and neutrons produced in
neutrino interactions are also important, because the proton tracks
are used to classify the neutrino event type.  Nucleon-nucleon
interactions modify the outgoing nucleon's momentum and
direction. Both elastic scattering and pion production are
considered. In order to simulate these interactions, a cascade model
is again used and the generated particles in the nucleus are tracked
using the same code as for the mesons.

No de-excitation gamma-ray from the carbon nucleus is simulated
when nuclear breakup occurs.

%% file: detector.tex
\section{Neutrino Detector}
\label{sec:detector}

The SciBooNE detector is located 100~m downstream from the beryllium
target on the axis of the beam.
The detector comprises three sub-detectors: a fully active and finely
segmented scintillator tracker (SciBar), an electromagnetic calorimeter
(EC), and a muon range detector (MRD).
\begin{figure}[bp]
  \begin{center}
    \includegraphics*[scale=0.7]{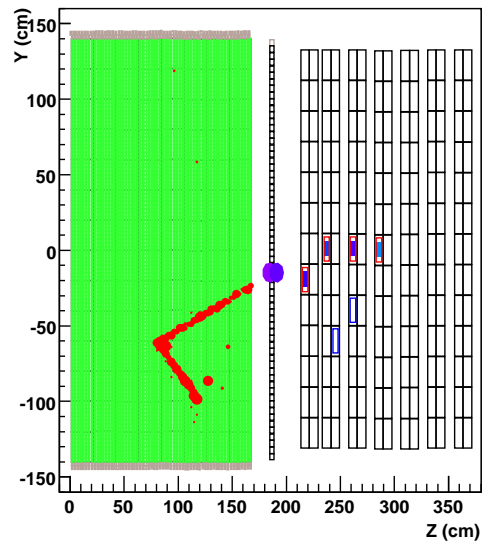}
    \caption{Event display of a typical muon neutrino charged current
      single charged pion event candidate in SciBooNE data.  
      The neutrino beam runs from left to right in this figure,
      encountering SciBar, the EC and MRD, in that order.
      The circles on SciBar and the EC indicate ADC hits for which the
      area of the circle is proportional to the energy deposition in that channel.
      Filled boxes in the MRD show ADC hits in time with the beam window.} 
    \label{fig:scibar_schematic}
  \end{center}
\end{figure}

\subsection{Detector Description}
\label{subsec:exp_appratus}

Fig.~\ref{fig:scibar_schematic} shows an event display of a typical
muon neutrino charged current single charged pion event
candidate.  Detector coordinates are shown
in the figure.  SciBooNE uses a right-handed Cartesian coordinate
system in which the $z$ axis is the beam direction and the $y$ axis is
the vertical upward direction.  The origin is located on the most
upstream surface of SciBar in the $z$ dimension, and at the center of
the SciBar scintillator plane in the $x$ and $y$ dimensions.  Since
each sub-detector is read out both vertically and horizontally, two
views are defined; the top view ($z$-$x$ projection) and the side view
($z$-$y$ projection).

The SciBar detector~\cite{Nitta:2004nt} is positioned upstream of the
other sub-detectors.  The primary role of SciBar is to reconstruct the
neutrino-nucleus interaction vertex and detect charged particles
produced by neutrino interactions.  Moreover, SciBar is capable of
particle identification based on deposited energy.  SciBar was
designed and built as a near detector for the K2K experiment.  After
K2K's completion, SciBar was relocated to the Fermilab BNB for SciBooNE.

SciBar consists of 14,336 extruded plastic scintillator strips which
serve as the target for the neutrino beam as well as the active detection
medium.  Originally produced by Fermilab, each strip has a dimension of
1.3 $\times$ 2.5 $\times$ 300~cm$^3$.  The scintillators are arranged
vertically and horizontally to construct a 3 $\times$ 3 $\times$
1.7~m$^3$ volume with a total mass of 15 tons.
Each strip is read out by a wavelength shifting (WLS) fiber attached
to a 64-channel multi-anode PMT.  Charge and timing information from
each PMT is recorded by front-end electronics boards (FEB) attached
directly to the PMT and a back-end VME module~\cite{Yoshida:2004mh}.
The FEB uses VA/TA ASICs; the VA handles charge information from the
PMT with a 32-channel preamplifier chip with a shaper and multiplexer,
while the TA provides timing information by taking the logical ``OR''
of 32 channels.  The charge and timing information are digitized by
ADC and multi-hit TDC modules on back-end electronics, and read out
through the VME bus.

The gains of all PMT channels were measured prior to installation in
K2K.  SciBar is equipped with a gain calibration system comprised of
LEDs to monitor and correct gain drift during the data taking; the
gain stability is monitored with precision better than 1\%.  Cosmic
ray data are also employed to calibrate the PMT gains and scintillator
light yield, including attenuation of the WLS-fibers.  These
calibration data, LED and cosmic, are taken between beam spills and
continuously monitored.  Calibration data verify the light yield was
stable within 1\% during operation.  The timing resolution for
minimum-ionizing particles was evaluated with cosmic ray data to be
1.6~ns.  The average light yield for minimum-ionizing particles is
approximately 20 photoelectrons per 1.3~cm path length, and the
typical pedestal width is below 0.3 photoelectron.  The hit finding
efficiency evaluated with cosmic ray data is more than 99.8\%.  The
minimum length of a reconstructable track is approximately 8~cm (three
layers hit in each view).  The track finding efficiency for single
tracks of 10~cm or longer is more than 99\%.

The EC is located just downstream of SciBar, and is designed to
measure the electron neutrino contamination in the beam and tag
photons from $\pi^0$ decay.  The EC is a ``spaghetti'' type
calorimeter comprised of 1~mm diameter scintillating fibers embedded
in lead foil.  The calorimeter is made of modules of dimensions 262
$\times$ 8 $\times$ 4~cm$^3$.  Each module is read out by two
green-extended 1~inch Hamamatsu PMTs per side, 256 PMTs total. The
modules were originally built for the CHORUS neutrino experiment at
CERN~\cite{Buontempo:1994yp} and later used in HARP and then K2K.  The
modules construct one vertical and one horizontal plane, and each
plane has 32 modules.  The EC has a thickness of 11 radiation lengths
along the beam direction.  The planes cover an active area of 2.7
$\times$ 2.6~m$^2$.

The charge information from each PMT is recorded.  A minimum ionizing
particle with a minimal path length deposits approximately 91~MeV in the EC.
The energy resolution for electrons was measured to be
14\%$/\sqrt{E \ {\rm (GeV)}}$ using a test beam~\cite{Buontempo:1994yp}.

The MRD is installed downstream of the EC and is designed to measure
the momentum of muons produced by charged-current neutrino
interactions.  The MRD was constructed for SciBooNE at Fermilab, primarily
out of parts recycled from past experiments.  It has 12 iron plates
with thickness of 5~cm which are sandwiched between planes of 6~mm
thick scintillation counters, 13 alternating horizontal and vertical
planes, which are read out via 2~inch PMTs from a variety of past
experiments; there are 362 PMTs total.  Each iron plate covers an area
of 274 $\times$ 305~cm$^2$.  The total mass of absorber material is
approximately 48~tons.  The MRD measures the momentum of muons up to
1.2~GeV/$c$ using the observed muon range.  Charge and timing
information from each PMT are recorded.  Hit finding efficiency was
continuously monitored using cosmic ray data taken between beam
spills; the average hit finding efficiency is 99\%.

SciBooNE has two global triggers, the beam trigger and the off-beam
trigger.  Two types of data are collected in one beam cycle, neutrino
data with the beam trigger and calibration data with the off-beam
trigger.  One cycle is about 2~sec which is defined by the accelerator
timing sequence.  The BNB receives one train of proton beam pulses per
cycle, with a maximum of 10 pulses in a row at 15~Hz.

A fast timing signal sent by the extraction magnet on BNB pulses establishes
a beam-trigger.  Once the beam trigger condition is set, all sub-detector
systems read out all channels irrespective of hit occupancy
(i.e. whether or not a neutrino interaction occurred), ensuring
unbiased neutrino data.

After the beam trigger turns off, the off-beam trigger condition is
automatically set and each sub-detector takes calibration data.  There
are three types of calibration data: pedestal, LED (only for SciBar)
and cosmic ray data.  The pedestal and LED data are collected once per
cycle.  For cosmic ray data, there are two independent trigger blocks:
SciBar/EC and MRD.  SciBar and the EC use a common cosmic ray trigger
which is generated using fast signals from the TA.  The MRD has its own
cosmic ray trigger which is also self-generated by discriminator outputs.
Both SciBar/EC and the MRD collect 20 cosmic ray triggers in a cycle.

\subsection{Detector simulation}
\label{subsec:detector_mc}

The GEANT4 framework is used for the detector simulation.  The Bertini
cascade model within GEANT4~\cite{Heikkinen:2003sc} is used to simulate
the interactions of hadronic particles with detector materials.
The detector simulation includes a detailed geometric model of the detector,
including the detector frame and experimental hall and soil, which is based on
survey measurements taken during detector construction.

In the detector simulation of SciBar, low level data parameters are
used as input to the simulation whenever possible.  The energy loss of
a charged particle in a single strip is simulated by GEANT, and this
energy scale is tuned using cosmic ray data.  Scintillator quenching
is simulated using Birk's law~\cite{Birks:1964aa} with a value of
Birk's constant, measured for K2K, of
0.0208~cm/MeV~\cite{Hasegawa:2006am}.  The energy deposited by a
charged particle is converted to photoelectrons using conversion
factors measured for each channel with cosmic muons.  The measured
light attenuation length of each fiber (approximately 350~cm on
average) is used in the simulation.  Crosstalk between nearby MA-PMT
channels is simulated using values measured in a test stand prior to
installation.  The number of photoelectrons is smeared by Poisson
statistics, and the single photoelectron resolution of the MA-PMT is
simulated.  To simulate the digitization of the PMT signal, the number
of photoelectrons is converted to ADC counts, and then electronics
noise and threshold effects of the TA are simulated.

TDC hit simulation includes light propagation delays in the WLS
fibers.  A logical OR of 32 MA-PMT channels is made for each TDC
channel, and the time of each hit is converted to TDC counts.
Multiple TDC hits in each channel are simulated.

In the EC detector simulation, true energy deposition in scintillating
fibers in the detector is converted to the number of photoelectrons
using a conversion factor which is measured for each channel with
cosmic-ray muons. The attenuation of light in the fiber is simulated
using the measured attenuation length value.  The number of photoelectrons
is smeared by Poisson statistics and by the PMT resolution,
and then converted to ADC counts.  The time-dependent ADC gain due to
the overshoot of the PMT signal is simulated based on a
measurement with cosmic muons. Electronics noise is also simulated.

For the detector simulation of the MRD, true energy deposition in each
scintillator is converted to ADC counts using the conversion factor
measured with cosmic muons.  The attenuation of light in the
scintillator as well as electronics noise are simulated.  Gaps between
scintillator counters in each plane, which cause inefficiency, are
included in the simulation. The time of energy deposition is digitized
and converted into TDC counts.

%% file: data.tex
\section{Data Summary}
\label{sec:data}

\begin{figure}[tbp]
  \begin{center}
    \includegraphics[keepaspectratio=true,height=58mm]{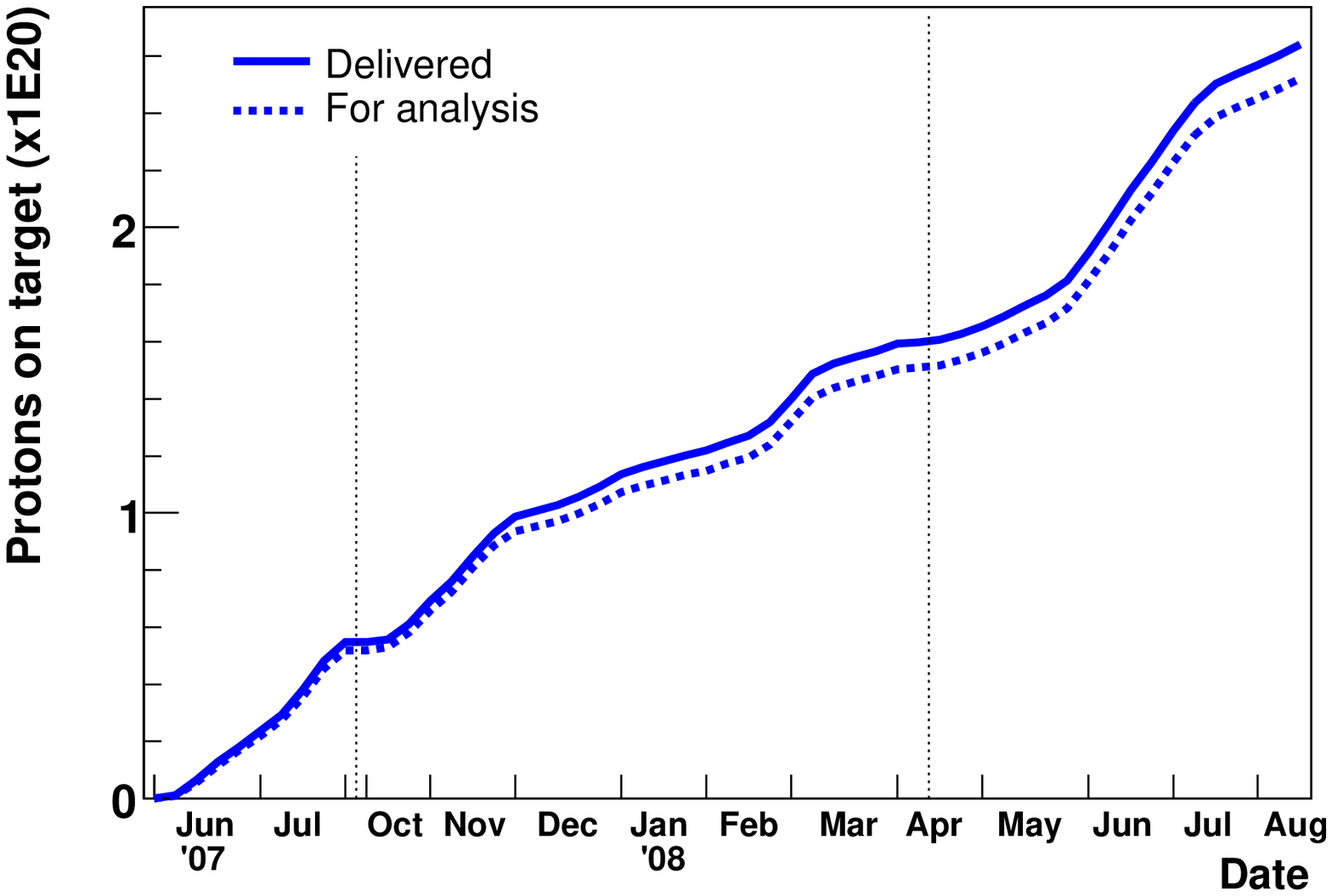}
    \includegraphics[keepaspectratio=true,height=58mm]{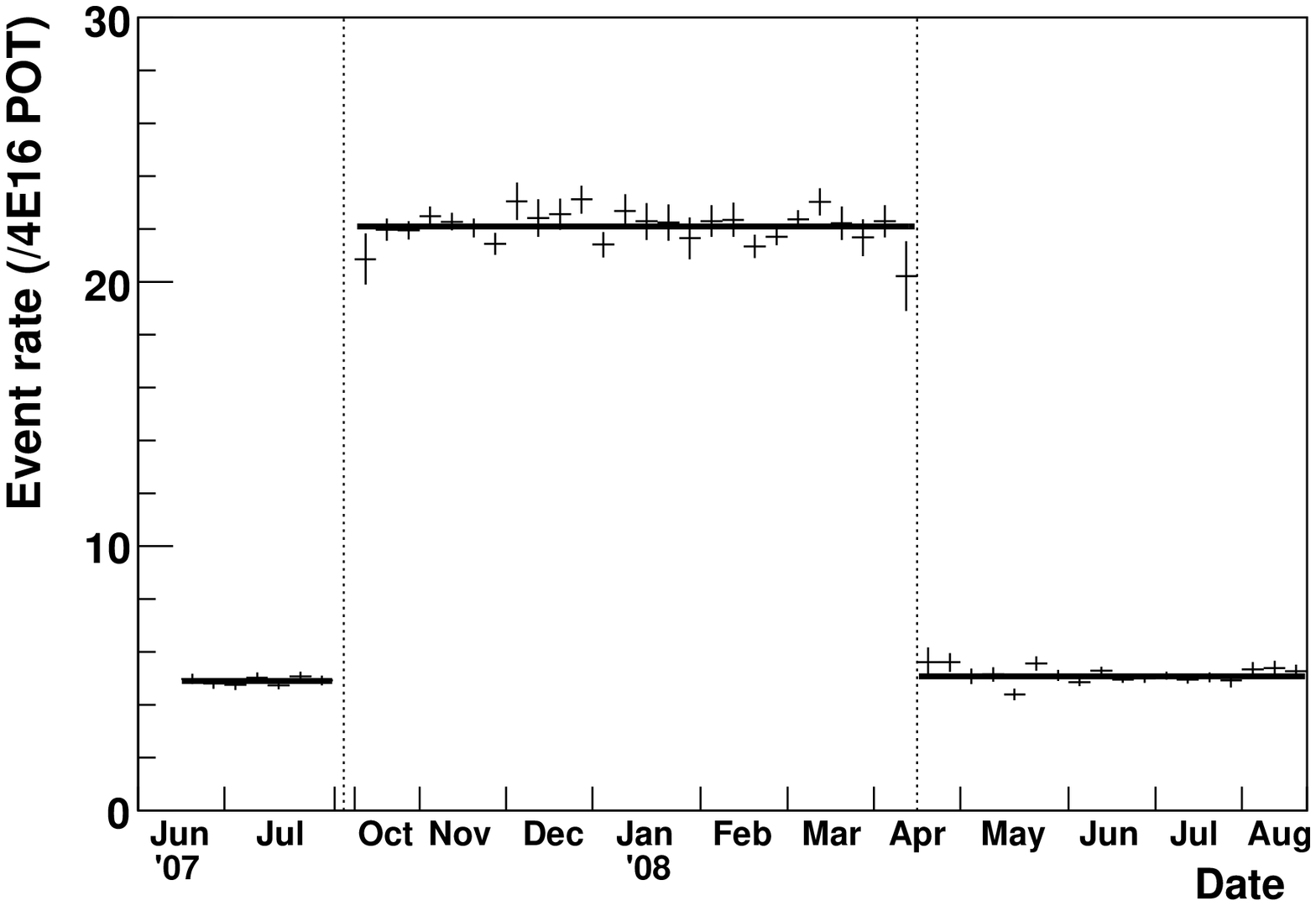}
  \end{center}
  \caption{Experimental performance.  In the top panel, the solid line
  shows the history of the accumulated number of protons on target and
  the dashed line shows the number of protons on target passing all data 
  quality cuts.  The bottom panel shows the number of charged current
  candidate events in SciBar normalized to the number of protons on
  target.  The event rate difference between neutrino and antineutrino 
  modes can be seen clearly.}
  \label{fig:potsummary.eps}
\end{figure}

The SciBooNE experiment took data from June~2007 until August~2008.
The data-taking is divided into three periods depending on the
polarity of the horn, as summarized in Table~\ref{table:run_summary}.
Fig.~\ref{fig:potsummary.eps}(top) shows a history of the accumulated
number of protons on target; the two curves show the total protons on
target for all events and the protons on target for events passing all
data quality cuts, described below.

\begin{table}[tbp]
 \caption{Summary of SciBooNE data-taking.  The table shows the number 
          of protons on target (POT) collected after application of data 
          quality cuts, as described in the text.}
 \label{table:run_summary}
 \begin{center}
  \begin{tabular}{llll}
    \hline \hline
     Run                  & Period                & POT                  \\
    \hline
     Run 1 (Antineutrino) & Jun. 2007 - Aug. 2007 & $0.52\times 10^{20}$ \\
     Run 2 (Neutrino)     & Oct. 2007 - Apr. 2008 & $0.99\times 10^{20}$ \\
     Run 3 (Antineutrino) & Apr. 2008 - Aug. 2008 & $1.01\times 10^{20}$ \\
    \hline \hline
  \end{tabular}
 \end{center}
\end{table}

In total, $2.64\times 10^{20}$ protons on target were delivered to the
beryllium target during the SciBooNE data run.  The beam datastream
(measuring, for example, magnet current settings, measured beam
intensity, measured peak horn current) is synchronized and merged with
the corresponding SciBooNE detector datastream, provided that the
spill time as measured by the beam instrumentation and by the detector
match within 10~ms of each other.

Only spills that satisfy certain beam quality cuts are used for
analysis. We require that the beam intensity is at least $0.1\times
10^{12}$ protons per spill, that the agreement between the two toroid
readouts along the beam-line is within 10\%, that the absolute peak horn
current is greater than 170 kA, and that the targeting efficiency is greater
than 95\%. Overall, beam quality cuts reject less than 1\% of the
total number of protons on target accumulated during the run. A
somewhat larger fraction of protons on target is rejected because of
detector dead time, yielding about a 95\% efficiency to satisfy all
(beam plus detector) data quality cuts.  After all beam and detector
quality cuts, $2.52\times 10^{20}$ protons on target are usable for physics
analyses.

In this analysis, the full neutrino data sample is used, corresponding
to $0.99\times 10^{20}$ protons on target satisfying all data quality
cuts, collected between October 2007 and April 2008. During that time,
all detector channels were operational on beam triggers.  The
experimental stability is demonstrated in
Fig.~\ref{fig:potsummary.eps}(bottom), which shows the number of
charged current event candidates per protons on target.  In this
figure, the event reconstruction is a simple $\chi^2$ track finder
which is used only for operations related studies, and not for the
analysis described in this paper.  The figure illustrates the
event rate difference between neutrino mode and antineutrino mode
running.

The antineutrino data sample collected before and after the neutrino
data-taking period is not considered in this analysis.

%% file: analysis.tex
\section{Event Reconstruction and Analysis}
\label{sec:analysis}

\subsection{Track Reconstruction}
The first step of the event reconstruction is to search for
two-dimensional tracks in each view of SciBar using a cellular
automaton algorithm \cite{Maesaka:2005aj}.  For tracking, the hit
threshold is set to two photoelectrons, corresponding to approximately
0.2~MeV.  Three dimensional tracks are reconstructed by matching
the timing and $z$-edges of the two dimensional tracks. The timing difference
between two two dimensional tracks is required to be less than 50 nsec, and the
$z$-edge difference must be less than 6.6~cm for upstream and
downstream edges. Reconstructed tracks are required to have at least
three-layer penetration, and therefore the minimum length of a
reconstructed track is 8~cm in the beam direction. According to the MC
simulation, 96\% of charged current interactions in SciBar are reconstructed to
have at least one track.

To identify charged current events, we look for events in which
at least one reconstructed track in SciBar is matched with a track
or hits in the MRD.  Such a track is defined as a SciBar-MRD matched
track. The most energetic SciBar-MRD matched track in any event is
considered as a muon candidate.
For matching a MRD track to a SciBar track, the upstream edge
of the MRD track is required to be on either one of the first two
layers of the MRD. 
The transverse distance between the two tracks at the first layer of
the MRD must be less than 30~cm.
The requirement on the difference between
track angles with respect to the beam direction is given by
$|\theta_{\rm MRD}~-~\theta_{\rm SB}|~<~\theta_{\rm max}$, where
$\theta_{\rm max}$ is a function of the length of the MRD track,
varying between 0.4~radian and
1.1~radians.  For track reconstruction in the MRD, at least two hit
layers in each view are needed, and thus this matching method is used
for tracks which penetrate at least three steel plates. If no MRD
track is found, we extrapolate the SciBar track to the MRD and search
for nearby contiguous hits in the MRD identifying a short muon track.
For matching MRD hits to a SciBar track, the MRD hit is required to be
within a cone with an aperture of $\pm$0.5~radian and a transverse
offset within 10~cm of the extrapolated SciBar track at the upstream
edge of the MRD. The timing difference between the SciBar track and
the track or hits in the MRD is required to be within 100 nsec.  The
matching criteria impose a muon momentum threshold of 350~MeV/$c$.

\subsection{Particle Identification}
The SciBar detector has the capability to distinguish protons from
muons and pions using $dE/dx$.
The particle identification variable, Muon Confidence Level (MuCL) is
calculated as follows.  The confidence level at each plane is first
defined as the fraction of events in the expected $dE/dx$ distribution
of muons above the observed value, $(dE/dx)_{\rm obs}$. The expected
$dE/dx$ distribution of muons is obtained by using cosmic-ray muons.
Each plane's confidence level is combined to form a total confidence
level, assuming the confidence level at each layer is independent. The
MuCL is calculated as
\begin{eqnarray}
  {\rm MuCL} = P \times \sum_{i=0}^{n-1} \frac{(-\ln P)^i}{i!}
\end{eqnarray}
where $n$ is the number of planes penetrated by the track,
$P = \prod_{i=1}^{n} {\rm CL}_{i}$, ${\rm CL}_{i}$ is the confidence level
at the $i$-th plane.

\begin{figure}[tbp]
  \begin{center}
    \includegraphics[keepaspectratio=true,height=40mm]{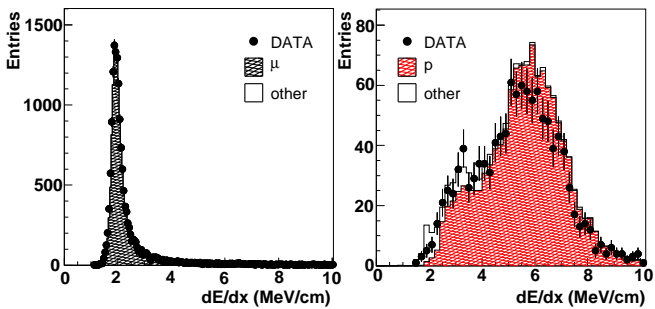}
  \end{center}
  \caption{(Color online) $dE/dx$ of muon enriched sample (left) and proton
  enriched sample (right).}
  \label{fig:dedx_mu_p.eps}
\end{figure}

Fig.~\ref{fig:dedx_mu_p.eps} shows the $dE/dx$ distributions of muon
and proton enriched samples. The predicted distributions of true muon
and proton tracks are shown as hatched histograms. To select muon
candidates for this study, we first select SciBar-MRD matched
tracks. According to the MC simulation, the sample is 94.7\% pure
muons with a small contamination of protons and charged pions. For
proton candidates, we select the second track in a charged current
quasi-elastic (CC-QE) scattering enriched sample made by cutting on a
kinematic variable described later.  The fraction of protons in the
sample is 92.1\%, estimated with the MC simulation. The contamination
of charged pions and muons are estimated to be 5.5\% and 1.6\%,
respectively.  Proton candidates are clearly separated from muon
candidates.

\begin{figure}[tbp]
  \begin{center}
    \includegraphics[keepaspectratio=true,height=40mm]{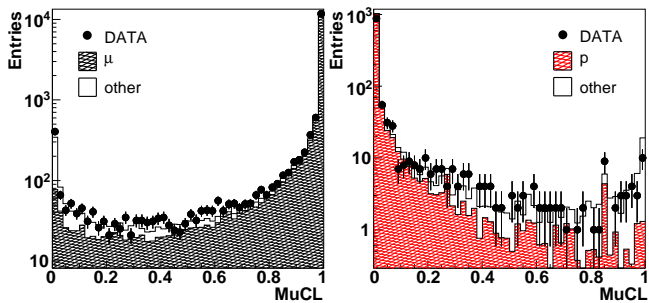}
  \end{center}
  \caption{(Color online) MuCL of muon enriched (left) and proton
  enriched (right) samples.}
  \label{fig:mucl_mu_p.eps}
\end{figure}

The MuCL distributions for the muon enriched sample and the proton
enriched sample are shown in Fig.~\ref{fig:mucl_mu_p.eps}. Tracks with
MuCL greater than 0.05 are considered muon-like (or pion-like) and the
others are classified as proton-like.  The probability of
misidentification is estimated to be 1.1\% for muons and 12\% for
protons, averaged over track length in the muon and proton enriched
samples.

\subsection{Charged Current Event Selection}

Events with at least one SciBar-MRD matched track are selected as charged current
event candidates.  We reject events with hits associated with the muon
candidate on the most upstream layer of SciBar to eliminate incoming
particles due to neutrino interactions in the upstream wall or
soil. The hit threshold for this veto cut is set to two
photoelectrons.  The neutrino interaction vertex is reconstructed as
the upstream edge of the muon track.  The vertex resolution is
approximately 0.5~cm in each dimension, estimated with the MC
simulation. We select events whose vertices are in the SciBar fiducial
volume, defined to be $\pm$130~cm in both the $x$ and $y$ dimensions,
and 2.62~cm$<z<$157.2~cm, a total mass of 10.6~tons. The background
contamination due to neutrino events which occur in the EC and MRD is
2.0\% and 0.5\%, respectively.  Finally, the time of the muon
candidate is required to be within a 2~$\mu$sec window around the beam
pulse. The cosmic-ray background contamination in the beam timing
window is only 0.5\%, estimated using a beam-off timing
window. According to the MC simulation, the selection efficiency and
purity of true $\nu_\mu$ charged current events are 27.9\% and 92.8\%,
respectively. Impurity comes from $\nu_\mu$ neutral current events (3.0\%),
$\overline{\nu}_\mu$ charged current events (1.6\%), and neutrino events which
occur in the EC/MRD (2.5\%).  The average neutrino beam energy for
true charged current events in the sample is 1.2~GeV.  This SciBar-MRD matched
sample is our standard charged current data set and defines the MC normalization,
i.e. the MC distributions are normalized to the number of SciBar-MRD
matched events in data.

Two sub-samples of the SciBar-MRD matched sample are further defined;
the MRD stopped sample and the MRD penetrated sample. Events with the
muon stopping in the MRD are classified as MRD stopped events,
in which we can measure the muon momentum.  Events with the muon exiting
from the downstream end of the MRD are defined as the MRD penetrated
sample, in which we can measure only a part of the muon momentum.  The
average neutrino beam energy for true charged current events in the MRD stopped and
MRD penetrated samples are 1.0~GeV and 2.0~GeV, respectively, enabling
a measurement of charged current coherent pion production at two different mean neutrino
energies.

The slopes of the muon angles with respect to the beam in the two
SciBar views are used to calculate the three dimensional muon angle
with respect to the beam ($\theta_\mu$).
The kinetic energy of the muon is calculated by
the range and expected energy deposition per unit length ($dE/dx$) in
SciBar, the EC and the MRD,
\begin{eqnarray}
    E_{\rm kin} &=& E^{\rm SB} + E^{\rm EC} + E^{\rm MRD} \nonumber \\
        &=& \left.\frac{dE}{dx}\right|_{\rm SB} L_{\rm SB}
          +\frac{\Delta E_0^{\rm EC}}{\cos \theta_\mu} + E^{\rm MRD}(L_{\rm MRD})
\end{eqnarray}
where $E^{\rm SB}$, $E^{\rm EC}$, and $E^{\rm MRD}$ are the energy
deposition in each detector.  $L_{\rm SB}$ and $L_{\rm MRD}$ are the
track length of the muon in SciBar and the range in the MRD,
respectively.  We set $dE/dx|_{\rm SB}$ to 2.04~MeV/cm, and $\Delta
E_0^{\rm EC}$, which is the energy deposited in the EC by a
horizontally traversing minimum ionizing particle, is set to 91~MeV,
estimated with the GEANT4 simulation.  $E^{\rm MRD}$ is calculated
from a range to energy lookup table based on the MC simulation.  For
muons stopping in the MRD, the average muon momentum and muon angular
resolutions are 50~MeV/$c$ and 0.9~degree, respectively. For muons
exiting the MRD, only a lower limit on muon momentum is obtained,
while the muon angle is determined with the same resolution as that of
stopping muons. The systematic uncertainty in the muon momentum scale
is estimated to be 2\% which is dominated by the difference among
various calculations of the range to energy lookup table.

\subsection{Event Classification}
\label{sec:event_classification}
\begin{figure}[tbp]
  \begin{center}
    \includegraphics[keepaspectratio=true,height=60mm]{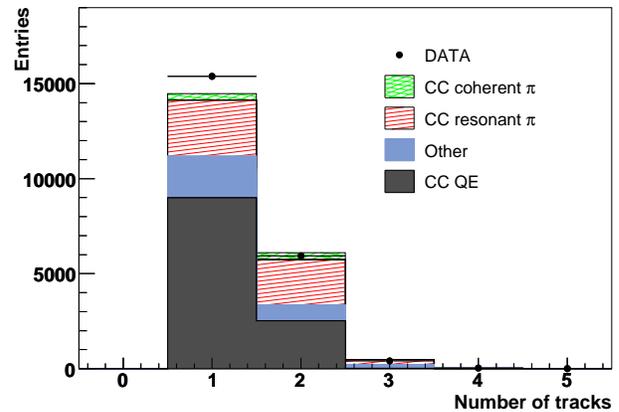}
  \end{center}
  \caption{(Color online) Number of vertex-matched tracks for the MRD stopped sample.
   The MC distribution shown here is before tuning.}
  \label{fig:ntracks.eps}
\end{figure}

\begin{figure}[tbp]
  \begin{center}
    \includegraphics[keepaspectratio=true,height=60mm]{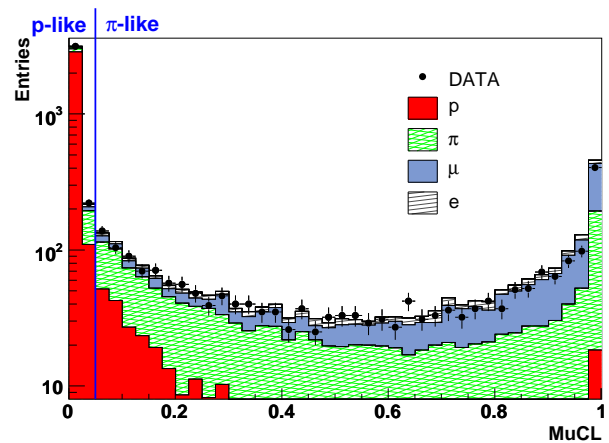}
  \end{center}
  \caption{(Color online) MuCL for the second track in the two-track sample.
  The MC distribution shown here is before tuning.}
  \label{fig:mucl.eps}
\end{figure}

The MRD stopped and MRD penetrated samples are further divided into sub-samples
with the same selection criteria.
Once the muon candidate and the neutrino interaction vertex are
reconstructed, we search for other tracks originating from the
vertex. For this purpose, the track edge distance is defined as the 3D
distance between the vertex and the closer edge of another
reconstructed track. Tracks whose edge distance is within 10~cm are
called vertex-matched tracks.  Fig.~\ref{fig:ntracks.eps} shows the
distribution of the number of tracks at the vertex for the MRD
stopped sample.  For the MC simulation, the contributions from
charged current coherent pion, charged current resonant pion, charged current quasi-elastic, and other interactions
are shown separately.
Most events are reconstructed as either one track or two track events.
The two track sample is further divided based on particle identification.  We
first require that the MuCL of the SciBar-MRD matched track is greater
than 0.05 to reject events with a proton penetrating into the MRD.
Then the second track in the event is classified as a pion-like or a
proton-like track with the same MuCL threshold.  Fig.~\ref{fig:mucl.eps}
shows the contributions to the second track from
true proton, pion, muon, and electron tracks as predicted
by the MC simulation.

\begin{figure}[tbp]
  \begin{center}
    \includegraphics[keepaspectratio=true,height=60mm]{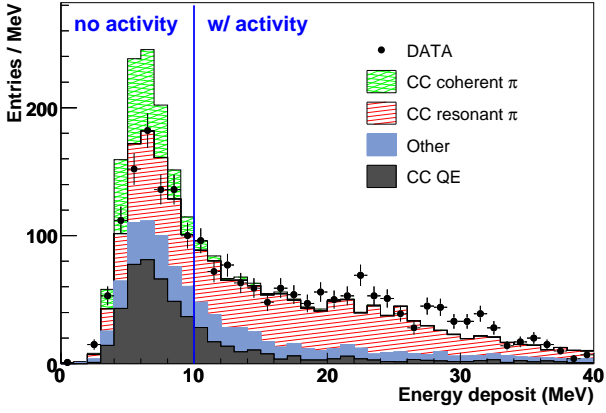}
  \end{center}
  \caption{(Color online) Maximum deposited energy in a strip around the vertex
  for the $\mu+\pi$ events. The MC distribution shown here is before tuning.}
  \label{fig:vtxedep.eps}
\end{figure}

In a charged current resonant pion event, $\nu p\to \mu^- p\pi^+$, the proton is
often not reconstructed due to its low energy, and such an event is
therefore identified as a two track $\mu+\pi$ event. To separate charged current
coherent pion events from charged current resonant pion events, additional protons
with momentum below the tracking threshold are instead detected by
their large energy deposition around the vertex, so-called vertex
activity.  We search for the maximum deposited energy in a strip
around the vertex, an area of $12.5 \ {\rm cm} \times 12.5 \ {\rm cm}$
in both views.  Fig.~\ref{fig:vtxedep.eps} shows the maximum energy
for $\mu+\pi$ events in the MRD stopped sample.
A peak around 6~MeV corresponds to the energy deposited in the strip containing the vertex 
by two minimum ionizing particles, and a high energy tail is mainly due to the low energy proton.
To simulate such protons, we consider re-interactions of nucleons in the nucleus as described
in Sec.~\ref{sec:intra_nuclear} as well as ones outside the nucleus
(described in Sec.~\ref{subsec:detector_mc}).
De-excitation gamma-rays from the carbon nucleus do not affect the distribution since most of
the gamma-rays first interact outside the vertex region.
Events with energy deposition greater than 10 MeV are considered to have activity at the
vertex. Charged current coherent pion candidates are extracted from the sample of
$\mu+\pi$ events without vertex activity. Four sub-samples, the
one track events, $\mu+p$ events, $\mu+\pi$ events with vertex activity,
and $\mu+\pi$ events without vertex activity in the MRD stopped sample
are used for constraining systematic uncertainties in the
simulation, described next.

\subsection{Tuning the Monte Carlo Simulation}
The MC simulation includes systematic uncertainties due to the detector response,
nuclear effects, neutrino interaction models, and neutrino beam
spectrum, and these uncertainties affect background estimation. The sources of
systematic uncertainty are summarized in Section~\ref{subsec:systematic}.
In order to constrain these uncertainties, the MC distributions of
the square of the four-momentum transfer ($Q^2$) are fitted to the distributions
of the four aforementioned data samples. The reconstructed $Q^2$ is calculated as
\begin{equation}
  Q^{2}_{\rm rec} = 2 E_{\nu}^{\rm rec} ( E_{\mu} - p_{\mu} \cos \theta_{\mu} ) - m_{\mu}^2
  \label{eq:q2rec}
\end{equation}
where $E_\nu^{\rm rec}$ is the reconstructed neutrino energy calculated by assuming
charged current quasi-elastic kinematics,
\begin{equation}
  E_{\nu}^{\rm rec} = \frac{1}{2}
  \frac{(m_p^2-m_\mu^2)-(m_n-V)^2+2E_\mu(m_n-V)}{(m_n-V)-E_\mu+p_\mu \cos \theta_\mu}
  \label{eq:enurec}
\end{equation}
where $m_p$ and $m_n$ are the mass of proton and neutron,
respectively, and $V$ is the nuclear potential, which is set to 27~MeV.
The one track events, $\mu+p$ events, and $\mu+\pi$ events with and without
vertex activity are fit simultaneously. Each $Q^2_{\rm rec}$
distribution is fit in bins of width 0.05~(GeV/$c$)$^2$ up to
1~(GeV/$c$)$^2$.

We introduce eight fitting parameters; the normalization factor of the
MRD stopped sample ($R_{\rm norm}$), the resonant pion scale factor
($R_{\rm res}$), the scale factor of other non-QE interactions
($R_{\rm other}$), the ratio of the number of two track events to the
number of one track events ($R_{\rm 2trk/1trk}$), the ratio of the
number of $\mu+p$ events to the number of $\mu+\pi$ events
($R_{p/\pi}$), the ratio of the number of low vertex activity
$\mu+\pi$ events to the number of high vertex activity $\mu+\pi$
events ($R_{\rm act}$), the muon momentum scale ($R_{\rm pscale}$),
and a charged current quasi-elastic Pauli-suppression parameter~$\kappa$.  All parameters are
ratios to nominal values in the MC simulation, i.e. all parameters
are set to 1 in the default MC simulation.

The parameters $R_{2trk/1trk}$, $R_{p/\pi}$, and $R_{\rm act}$
represent possible migrations between subsamples due to systematic
uncertainties.  The parameter $R_{\rm pscale}$ changes the scale of
the reconstructed muon momentum for the MC simulation. The parameter
$\kappa$, which was first introduced by MiniBooNE \cite{:2007ru},
controls the strength of Pauli-blocking and thus
suppresses low $Q^2$ charged current quasi-elastic events.  We employ this parameter in the
fitting because a deficit of data is found at low $Q^2$ in the one track
sample where the charged current quasi-elastic interaction is dominant.

The $\chi^2$ function to be minimized is given by:
\begin{eqnarray}
 \chi^2 = \chi^2_{\rm dist} + \chi^2_{\rm sys}.
\end{eqnarray}
The term $\chi^2_{\rm dist}$ is calculated using a binned likelihood defined as~\cite{Baker:1983tu}:
\begin{eqnarray}
 \chi^2_{\rm dist} &=&  -2\sum_{i,\ j}
 \ln \frac{P(N_{ij}^{\rm obs};N_{ij}^{\rm exp})}{P(N_{ij}^{\rm obs};N_{ij}^{\rm obs})} \nonumber \\
 &=& 2 \sum_{i,\ j} \left( N_{ij}^{\rm exp}-N_{ij}^{\rm obs}+N_{ij}^{\rm obs}
 \times \ln \frac{N_{ij}^{\rm obs}}{N_{ij}^{\rm exp}} \right)
\end{eqnarray}
where $P(n,\nu)=\nu^n e^{-\nu}/n!$ is the Poisson probability of finding $n$ events with
a expectation value $\nu$,
$N_{ij}^{\rm obs}$ and $N_{ij}^{\rm exp}$ are the observed and expected
number of events in the $i$-th $Q^2$ bin in subsample $j$ ($j=$one track, $\mu+p$,
$\mu+\pi$ with high and low vertex activity), respectively.
The expected number of events for each sample is given by:
\begin{eqnarray}
  N_{i,\ {\rm 1trk}}^{\rm exp}
  &=& R_{\rm norm} \nonumber \\
  & & \cdot \left[ n_{i, {\rm 1trk}}^{\rm QE}+R_{\rm res} n_{i, {\rm 1trk}}^{\rm res}
  +R_{\rm other} n_{i, {\rm 1trk}}^{\rm other} \right] \\
  N_{i,\ \mu p}^{\rm exp}
  &=& R_{\rm norm} \cdot R_{\rm 2trk/1trk} \cdot R_{p/\pi} \nonumber \\
  & & \cdot \left[ n_{i,\mu p}^{\rm QE}+R_{\rm res} n_{i,\mu p}^{\rm res}
  +R_{\rm other} n_{i,\mu p}^{\rm other} \right] \\
  N_{i,\ \mu\pi {\rm H}}^{\rm exp} 
  &=& R_{\rm norm} \cdot R_{\rm 2trk/1trk} \nonumber \\
  & & \cdot \left[ n_{i,\mu\pi {\rm H}}^{\rm QE}+R_{\rm res} n_{i,\mu\pi {\rm H}}^{\rm res}
  +R_{\rm other} n_{i,\mu\pi {\rm H}}^{\rm other} \right]
\end{eqnarray}
\begin{eqnarray}
  N_{i,\ \mu\pi {\rm L}}^{\rm exp} 
  &=& R_{\rm norm} \cdot R_{\rm 2trk/1trk} \cdot R_{\rm act} \nonumber \\
  & & \cdot \left[ n_{i,\mu\pi {\rm L}}^{\rm QE}+R_{\rm res} n_{i,\mu\pi {\rm L}}^{\rm res}
  +R_{\rm other} n_{i,\mu\pi {\rm L}}^{\rm other} \right]
  \label{eq:num_mupi} 
\end{eqnarray}
where $n_{i,\ j}^{\rm QE}$, $n_{i,\ j}^{\rm res}$, $n_{i,\ j}^{\rm other}$
are the number of charged current quasi-elastic, charged current resonant pion, and other events
in each bin in each subsample, respectively.
$R_{\rm pscale}$ and $\kappa$ do not appear explicitly in these equations,
but $R_{\rm pscale}$ causes migration between $Q^2$ bins and $\kappa$ changes
$n_{i,\ j}^{\rm QE}$.

The term $\chi^2_{\rm sys}$, added to constrain systematic parameters, is 
calculated as:
\begin{eqnarray}
 \chi^2_{\rm sys} = (\bm{P_{sys}}-\bm{P_0}) \bm{V}^{-1} (\bm{P_{sys}}-\bm{P_0})
\end{eqnarray}
where $\bm{P_{sys}}$ represents the set of systematic parameters and
$\bm{P_0}$ is the set of parameter values before fitting, expressed as:
\begin{eqnarray}
  \bm{P_{sys}} = \left(
    \begin{array}{c}
      R_{\rm res} \\
      R_{\rm 2trk/1trk} \\
      R_{p/\pi} \\
      R_{\rm pscale}
    \end{array}
  \right)
  \quad , \quad
  \bm{P_0} = \left(
   \begin{array}{c}
      1 \\
      1 \\
      1 \\
      1
   \end{array}
 \right). 
\end{eqnarray}
$\bm{V}$ is a covariance matrix estimated by considering the possible variations
due to systematic uncertainties in the detector responses, nuclear effects,
neutrino interaction models, and neutrino beam spectrum.
We prepare several MC event sets by changing each underlying physics parameter, 
i.e. the source of systematic uncertainty, by $\pm 1 \sigma$. The covariance 
between two systematic parameters $p_i$ and $p_j$ is calculated as:
\begin{eqnarray}
  V_{ij}\equiv {\rm cov}[p_i,p_j] 
  = \sum_{\rm source} \frac{\Delta p_i \Delta p_j |_+ + \Delta p_i \Delta p_j |_-}{2}
\end{eqnarray}
where $\Delta p_i \Delta p_j |_{+(-)}$ is the product of variations of two parameters
when the underlying physics parameter is increased (decreased) by the size of its 
uncertainty. The covariance matrix is estimated to be:
\begin{eqnarray}
  \bm{V} = \left(
    \begin{array}{cccc}
     \ (0.20)^2 &  -(0.09)^2 &  +(0.10)^2 & 0 \\
     -(0.09)^2  & \ (0.09)^2 &  -(0.07)^2 & 0 \\
     +(0.10)^2  &  -(0.07)^2 & \ (0.15)^2 & 0 \\
             0  &          0 &          0 & (0.02)^2
    \end{array}
  \right).
\end{eqnarray}
$R_{\rm norm}$, $R_{\rm other}$, $R_{\rm act}$, and $\kappa$ are unconstrained in the fit.

Events with $Q^2_{\rm rec}<0.10 \ ({\rm GeV}/c)^2$ in the $\mu+\pi$
sample with low activity are not included in the fit to avoid charged current
coherent pion signal events.  A data excess is observed in the region
with $Q^2_{\rm rec}<0.15
\ ({\rm GeV}/c)^2$ in the $\mu+p$ sample.
Further investigation reveals that the second track in the excess
events is emitted at a relatively large angle with respect to the beam
direction and has large $dE/dx$, thus the events have an
additional large energy deposition at the vertex. Each of these events
seems to have a muon and a proton with additional activity, and
therefore the excess is not expected to affect the charged current coherent pion
analysis.  A possible candidate for the excess is charged current resonant
pion production where the pion is absorbed in the nucleus. In such an
event, two or more additional nucleons should be emitted after the
pion is absorbed, which is currently not simulated.  The excess cannot
be explained with the introduced fitting parameters, and therefore the
region is not used in the fit.

\begin{figure}[htbp]
  \begin{center}
    \includegraphics[keepaspectratio=true,height=53mm]{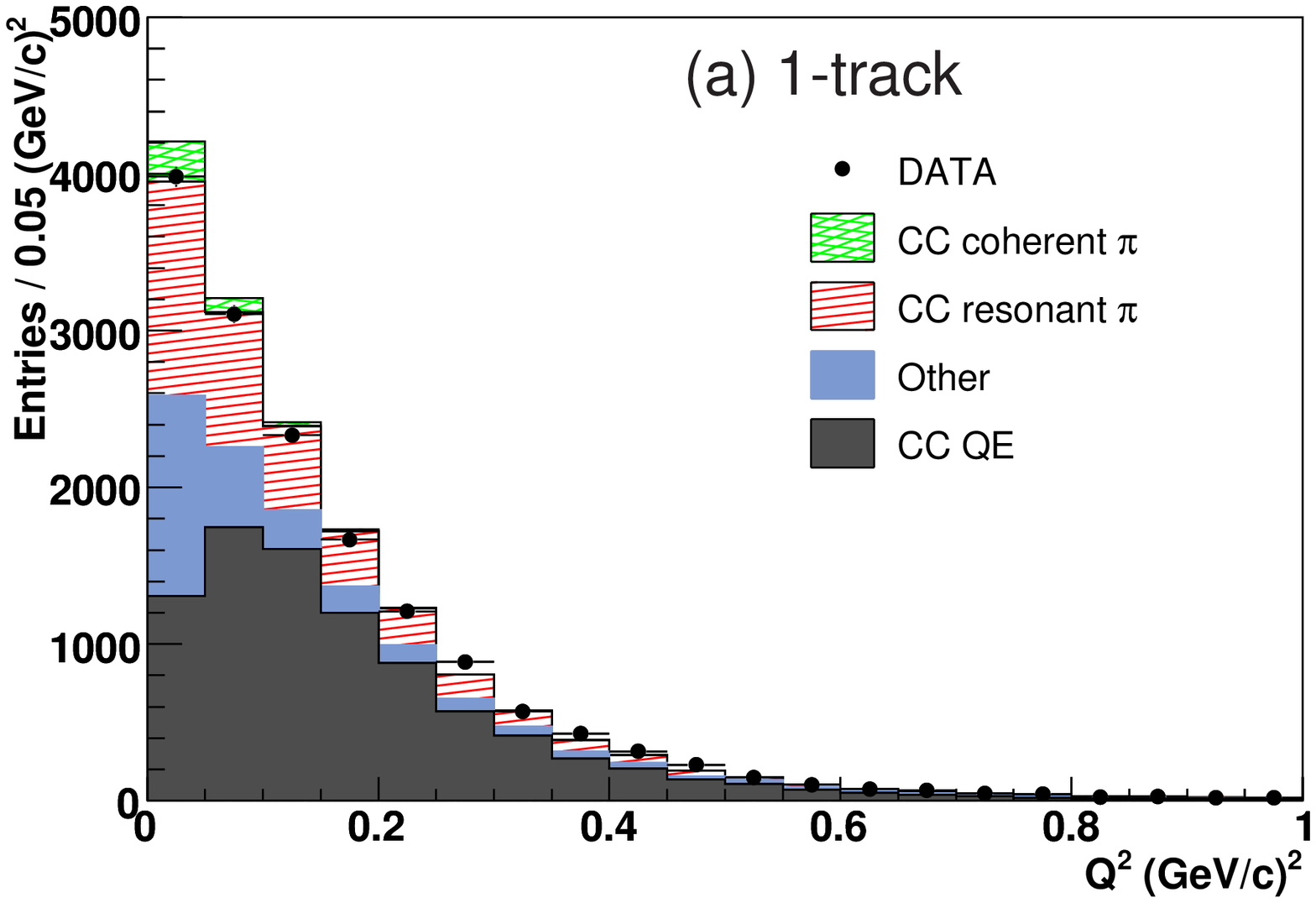}
    \includegraphics[keepaspectratio=true,height=53mm]{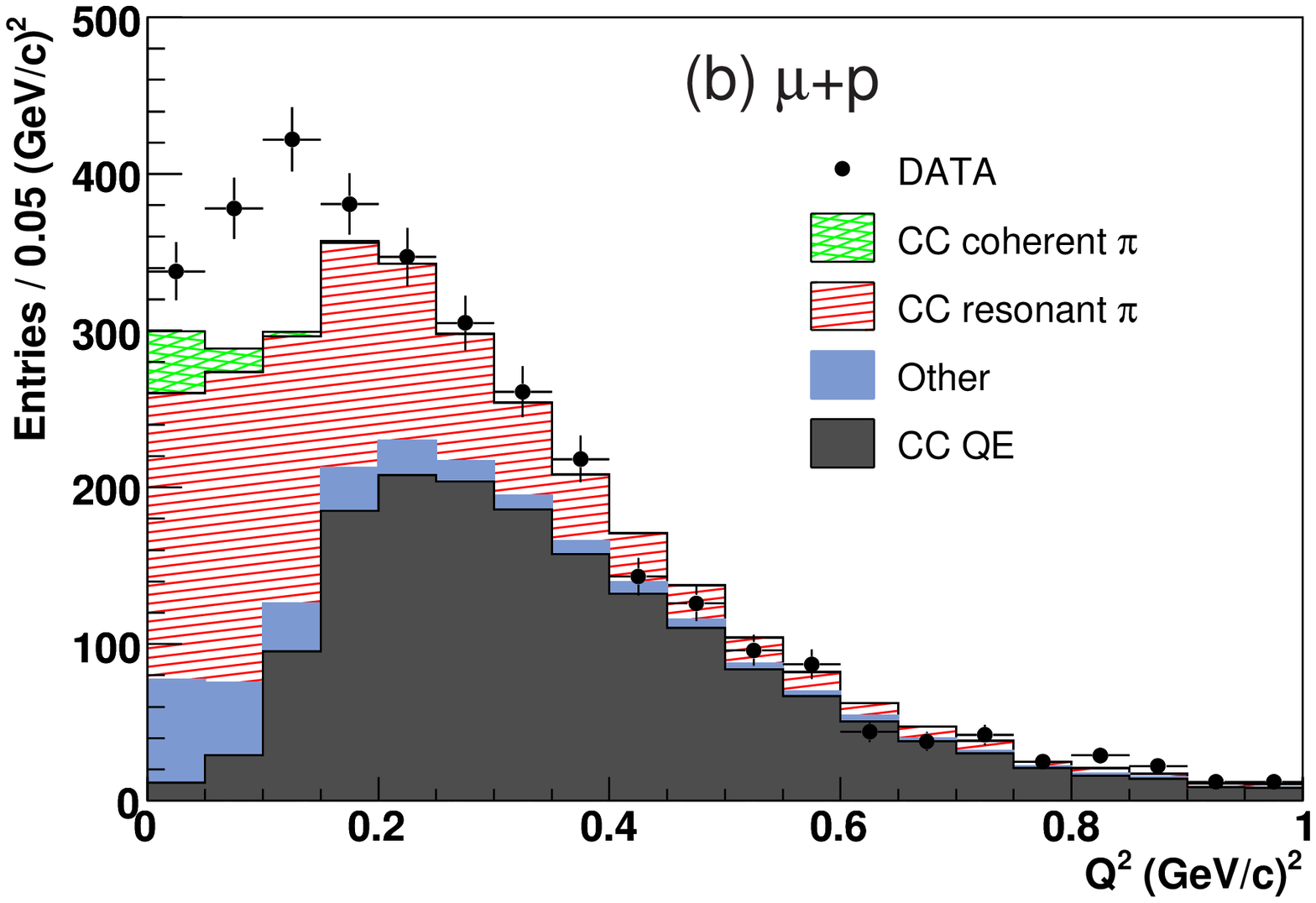}
    \includegraphics[keepaspectratio=true,height=53mm]{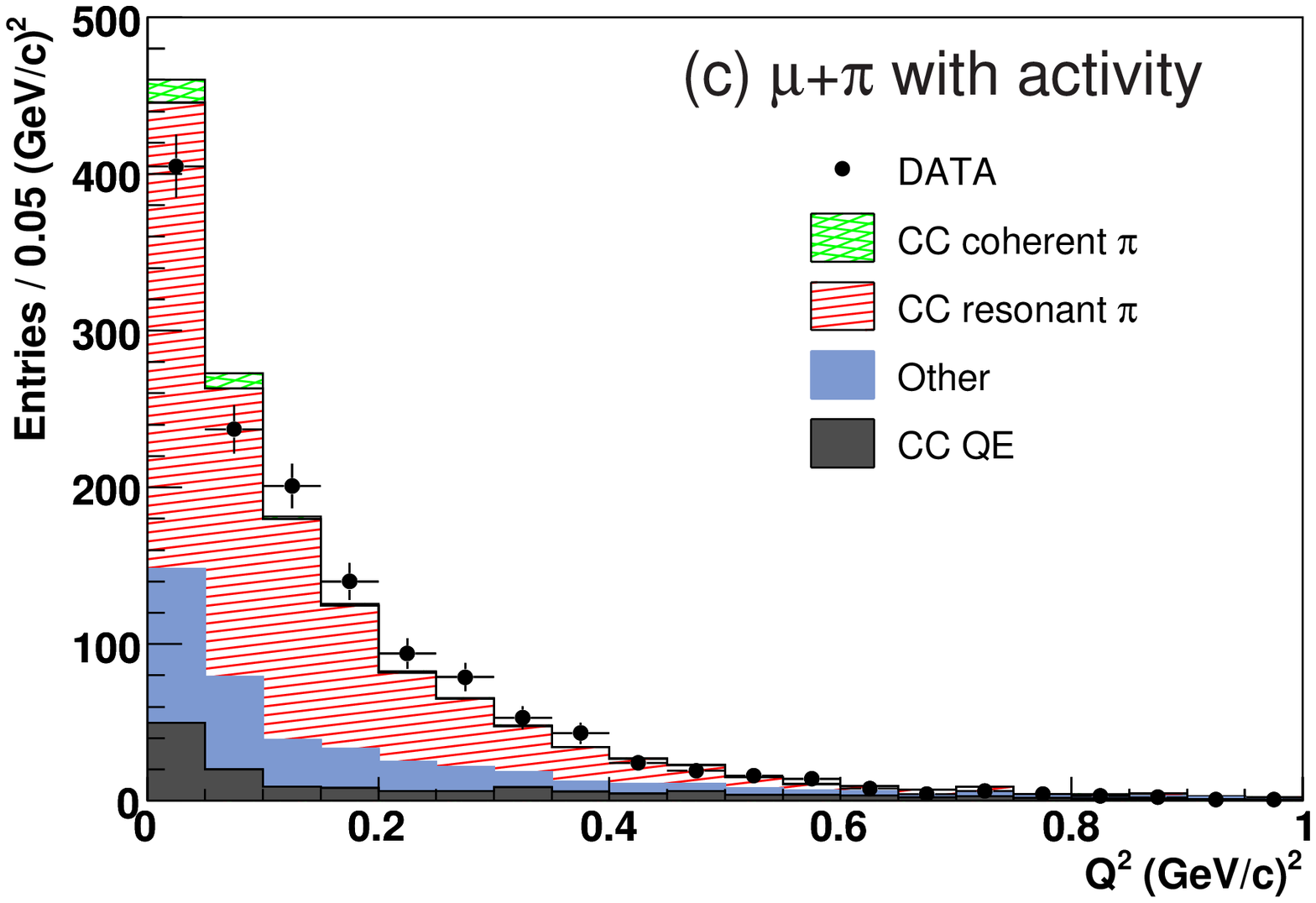}
    \includegraphics[keepaspectratio=true,height=53mm]{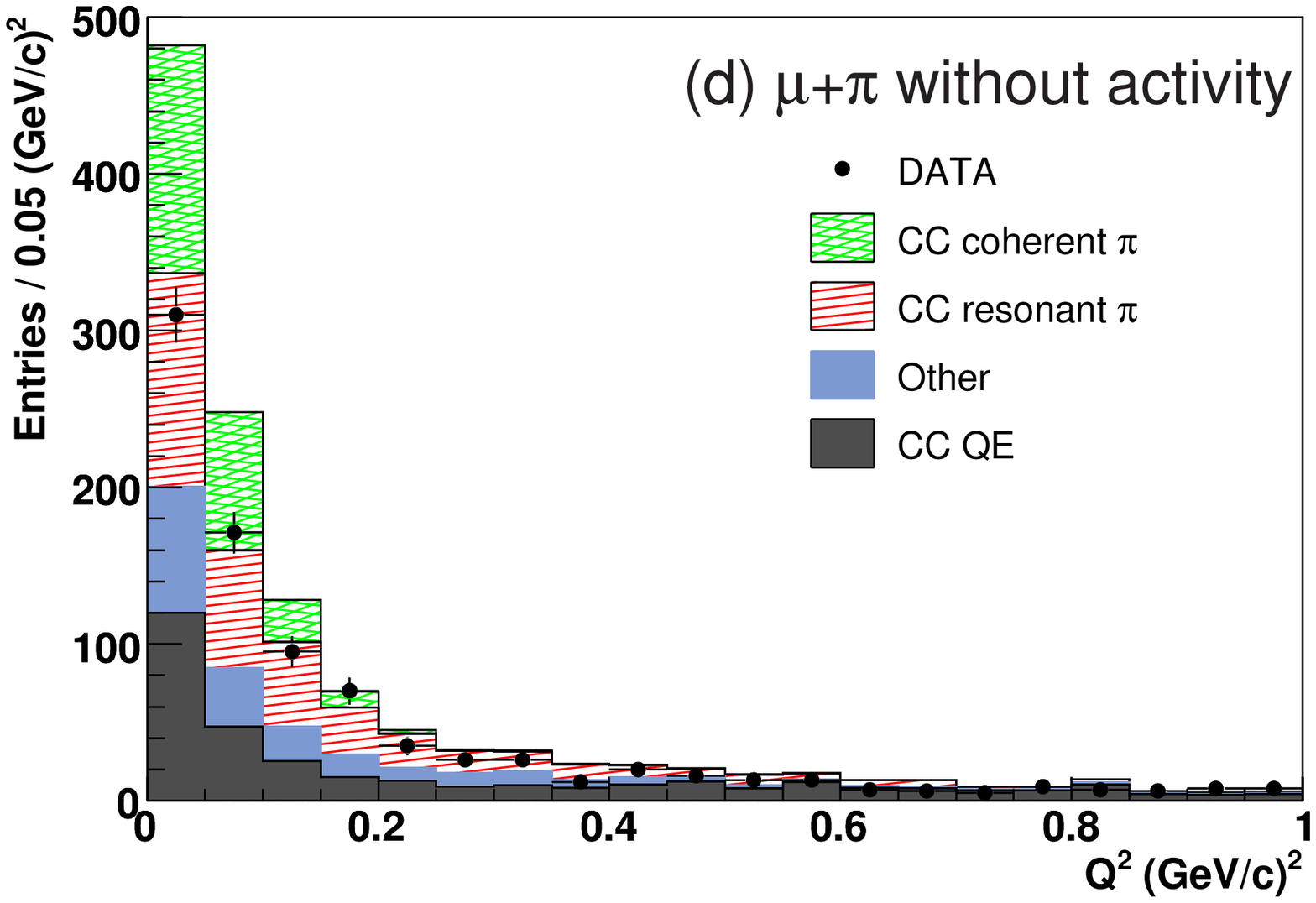}
  \end{center}
  \caption{(Color online) Reconstructed $Q^2$ after fitting for (a) the one track,
  (b) $\mu+p$, (c) $\mu+\pi$ with activity, and (d) $\mu+\pi$ without activity samples.}
  \label{fig:q2rec_after.eps}
\end{figure}

Fig.~\ref{fig:q2rec_after.eps} shows reconstructed $Q^2$ after the
fitting for the one track, $\mu+p$, and $\mu+\pi$ events with and without
vertex activity.  The best fit values and errors of the fit
parameters are summarized in Table~\ref{table:best_fit}.  These same
fit parameters are also applied to the MRD penetrating sample.  The
$\chi^{2}/$d.o.f before the fit is $473/75=6.31$.  The
$\chi^{2}/$d.o.f after the fit is $117/67=1.75$. Even after fitting,
the reduced $\chi^2$ is relatively large, which indicates that the
introduced parameters are not sufficient in fully reproducing the data. To take
into account the incompleteness of our simulation, we enlarge the errors
on the fitting parameters by a factor of $\sqrt{\chi^2/{\rm d.o.f}}$.

\begin{table}[htbp]
 \caption{Best fit values and errors of the fitting parameters}
 \label{table:best_fit}
 \begin{center}
  \begin{tabular}{lrrr}
    \hline \hline
    Parameter \quad & \quad Value & \quad Error \\
    \hline
    $R_{\rm norm}$      & 1.103 & 0.029 \\
    $R_{\rm 2trk/1trk}$ & 0.865 & 0.035 \\
    $R_{p/\pi}$         & 0.899 & 0.038 \\
    $R_{\rm act}$       & 0.983 & 0.055 \\
    $R_{\rm pscale}$    & 1.033 & 0.002 \\
    $R_{\rm res}$       & 1.211 & 0.133 \\
    $R_{\rm other}$     & 1.270 & 0.148 \\
    $\kappa$            & 1.019 & 0.004 \\
    \hline \hline
  \end{tabular}
 \end{center}
\end{table}

\subsection{Charged Current Coherent Pion Event Selection}
Charged current coherent pion candidates are extracted from both the MRD stopped and MRD penetrated
samples with the same selection criteria. In this section, we describe the event
selection for the MRD stopped sample. The event selection for the MRD penetrated
sample is summarized later.

\begin{figure}[tbp]
  \begin{center}
    \includegraphics[keepaspectratio=true,height=60mm]{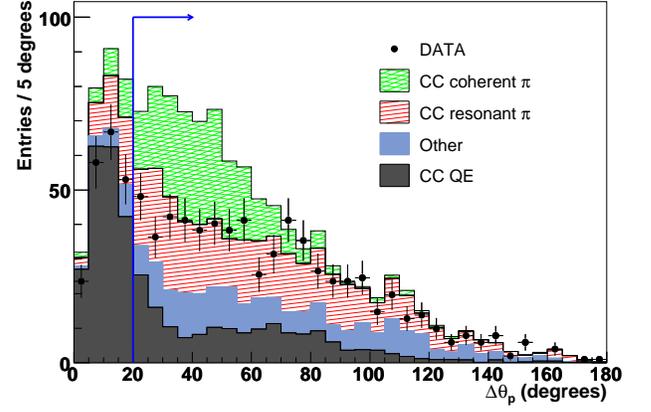}
  \end{center}
  \caption{(Color online) $\Delta \theta_p$ for the $\mu+\pi$ events in the
  MRD stopped sample after fitting.}
  \label{fig:dthetap.eps}
\end{figure}

After selecting $\mu+\pi$ events which do not have vertex activity,
the sample still contains charged current quasi-elastic events in which a proton is
misidentified as a minimum ionizing track. We reduce this charged current quasi-elastic background by
making use of kinematic information in the event.  Since the charged current quasi-elastic
interaction is a two-body interaction, one can predict the proton
direction from the measured muon momentum $p_\mu$ and muon angle
$\theta_\mu$;
\begin{equation}
  \vec{p}_p = (-p_{\mu x},-p_{\mu y},E_\nu^{\rm rec}-p_\mu \cos \theta_\mu)
  \label{eq:expected_proton_direction}
\end{equation}
where $p_{\mu x}$ and $p_{\mu y}$ are the projected muon momentum in the $x$ and $y$
dimension, respectively. $E_\nu^{\rm rec}$ is the reconstructed neutrino energy given by
Equation~\ref{eq:enurec}.
For each two-track event, we define an angle called $\Delta\theta_p$ as the angle between
the expected proton track direction given by Equation~\ref{eq:expected_proton_direction} and
the observed second track direction.
Fig.~\ref{fig:dthetap.eps} shows the $\Delta\theta_p$ distribution for $\mu+\pi$ events
in the MRD stopped sample. Events with $\Delta\theta_p$ larger than 20 degrees are selected.
With this selection, 48\% of charged current quasi-elastic events in the $\mu+\pi$ sample are
rejected, while 91\% of charged current coherent pion events pass the cut according to the MC simulation.

\begin{figure}[tbp]
  \begin{center}
    \includegraphics[keepaspectratio=true,height=60mm]{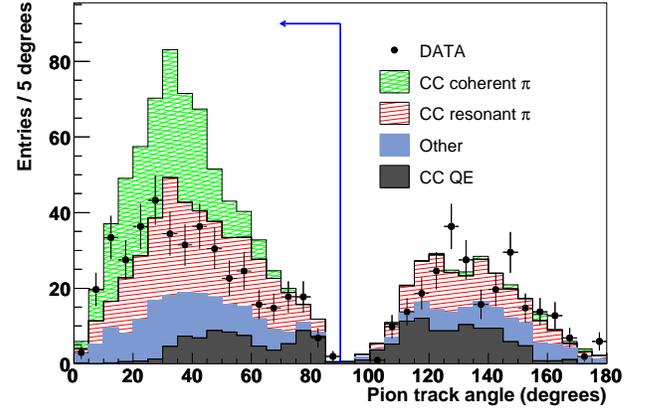}
  \end{center}
  \caption{(Color online) Track angle of the pion candidate with respect to
  the beam direction for the $\mu+\pi$ events after the charged current quasi-elastic rejection after
  fitting.}
  \label{fig:theta2nd.eps}
\end{figure}

Further selections are applied in order to separate charged current coherent pion
events from charged current resonant pion events which are the dominant backgrounds
for this analysis.  Fig.~\ref{fig:theta2nd.eps} shows the angular
distribution of pion candidates with respect to the beam direction.
In the case of charged current coherent pion events, both the muon and pion tracks
are directed forward. Events in which the track angle of the pion
candidate with respect to the beam direction is less than 90~degrees
are selected.

\begin{figure}[tbp]
  \begin{center}
    \includegraphics[keepaspectratio=true,height=60mm]{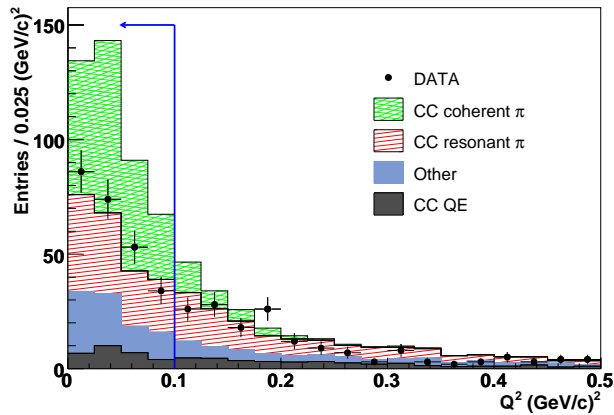}
  \end{center}
  \caption{(Color online) Reconstructed $Q^2$ for the $\mu+\pi$ events in the MRD
  stopped sample after the pion track direction cut and after fitting.}
  \label{fig:q2rec_mrdstop.eps}
\end{figure}

Fig.~\ref{fig:q2rec_mrdstop.eps} shows the reconstructed $Q^2$
distribution for the $\mu+\pi$ events after the pion track direction
cut. Although a charged current quasi-elastic interaction is assumed, the $Q^2$ of charged current coherent
pion events is reconstructed with a resolution of 0.016~(GeV/$c$)$^2$
and a shift of -0.024~(GeV/$c$)$^2$ according to the MC simulation.
Finally, events with reconstructed $Q^2$ less than 0.1 (GeV/$c$)$^2$
are selected.  The charged current coherent pion event selection is summarized in
Table~\ref{table:selection_mrdstop}.  In the signal region, 247 charged current
coherent pion candidates are observed, while the expected number of
background events is 228$\pm$12. The error comes from the errors on
the fitting parameters summarized in Table~\ref{table:best_fit}.  The
background in the final sample is dominated by charged current resonant pion
production.  The ``other'' background is comprised of 50\% charged current DIS,
32\% neutral current, and 18\% $\overline{\nu}_\mu$ events.  The selection
efficiency for the signal is estimated to be 10.4\%.

\begin{table}[tbp]
 \caption{Event selection summary for the MRD stopped charged current coherent pion sample.}
 \label{table:selection_mrdstop}
 \begin{center}
  \begin{tabular}{lrrrrr}
    \hline \hline
    Event selection             & DATA   & \multicolumn{2}{c}{MC} & Coherent $\pi$ \\
                                &        & Signal & B.G. & Efficiency \\
    \hline
    Generated in SciBar fid.vol.&        & 1,939 & 156,766 & 100\%   \\
    SciBar-MRD matched          & 30,337 &   978 &  29,359 & 50.4\% \\
    \hline
    MRD stopped                 & 21,762 &   715 &  20,437 & 36.9\% \\
    2 track                     &  5,939 &   358 &   6,073 & 18.5\% \\
    Particle ID ($\mu+\pi$)     &  2,255 &   292 &   2,336 & 15.1\% \\
    Vertex activity cut         &    887 &   264 &     961 & 13.6\% \\
    CC-QE rejection             &    682 &   241 &     709 & 12.4\% \\
    Pion track direction cut    &    425 &   233 &     451 & 12.0\% \\
    Reconstructed $Q^2$ cut     &    247 &   201 &     228 & 10.4\% \\    
    \hline \hline
  \end{tabular}
 \end{center}
\end{table}

\subsection{MRD penetrated Charged Current Coherent Pion Events}
\begin{figure}[tbp]
  \begin{center}
    \includegraphics[keepaspectratio=true,height=60mm]{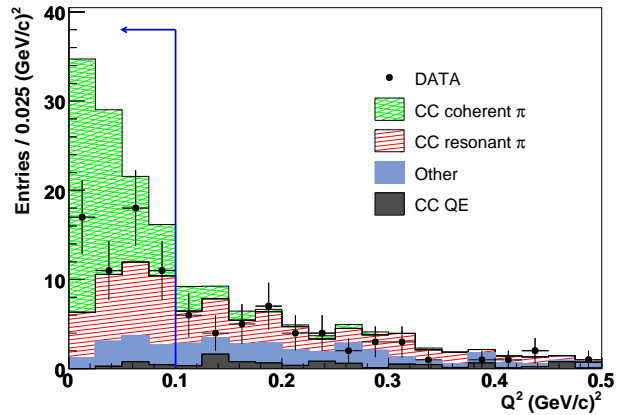}
  \end{center}
  \caption{(Color online) Reconstructed $Q^2$ for the $\mu+\pi$ events in the MRD
  penetrated sample after the pion track direction cut after fitting.}
  \label{fig:q2rec_mrdpenetrate.eps}
\end{figure}

The same selection is applied to the MRD penetrated sample to extract
charged current coherent pion candidates at higher energy. Fig.~\ref{fig:q2rec_mrdpenetrate.eps}
shows the reconstructed $Q^2$ distribution of the MRD penetrated charged current
coherent pion sample.  The reconstructed $Q^2$ and $E_\nu$ for the MRD
penetrated sample are calculated from muon angle and
partially-reconstructed muon energy, using Equation~\ref{eq:q2rec} and
Equation~\ref{eq:enurec}, respectively.  Although only a part of the
muon energy is observed, the $Q^2$ reconstruction performance is essentially
same because of the small muon angle. The event selection is
summarized in Table~\ref{table:selection_mrdpenetrate}.  In the signal
region, 57 charged current coherent pion candidates are observed, while the expected
number of background events is 40$\pm$2.2.  The background in the
final sample is dominated by charged current resonant pion production.  The
``other'' background is comprised of 75\% charged current DIS, and 25\%
$\overline{\nu}_\mu$ events.  The selection efficiency for the signal
is estimated to be 3.1\%.

\begin{table}[tbp]
 \caption{Event selection summary of MRD penetrated charged current coherent pion sample.}
 \label{table:selection_mrdpenetrate}
 \begin{center}
  \begin{tabular}{lrrrrr}
    \hline \hline
    Event selection             & DATA   & \multicolumn{2}{c}{MC} & Coherent $\pi$ \\
                                &        & Signal & B.G. & Efficiency \\
    \hline
    Generated in SciBar fid.vol.&        & 1,939 & 156,766 & 100\%   \\
    SciBar-MRD matched          & 30,337 &   978 &  29,359 & 50.4\% \\
    \hline
    MRD penetrated              &  3,712 &   177 &   4,375 &  9.1\% \\
    2 track                     &  1,029 &    92 &   1,304 &  4.7\% \\
    Particle ID ($\mu+\pi$)     &    418 &    78 &     474 &  4.0\% \\
    Vertex activity cut         &    167 &    71 &     186 &  3.6\% \\
    CC-QE rejection             &    134 &    67 &     135 &  3.5\% \\
    Pion track direction cut    &    107 &    66 &     109 &  3.4\% \\
    Reconstructed $Q^2$ cut     &     57 &    60 &      40 &  3.1\% \\    
    \hline \hline
  \end{tabular}
 \end{center}
\end{table}

%% file: results.tex
\section{Results}
\label{sec:results}

\subsection{Cross Section Ratio}
\subsubsection{MRD stopped charged current coherent pion sample}
\begin{figure}[htbp]
  \begin{center}
    \includegraphics[keepaspectratio=true,height=50mm]{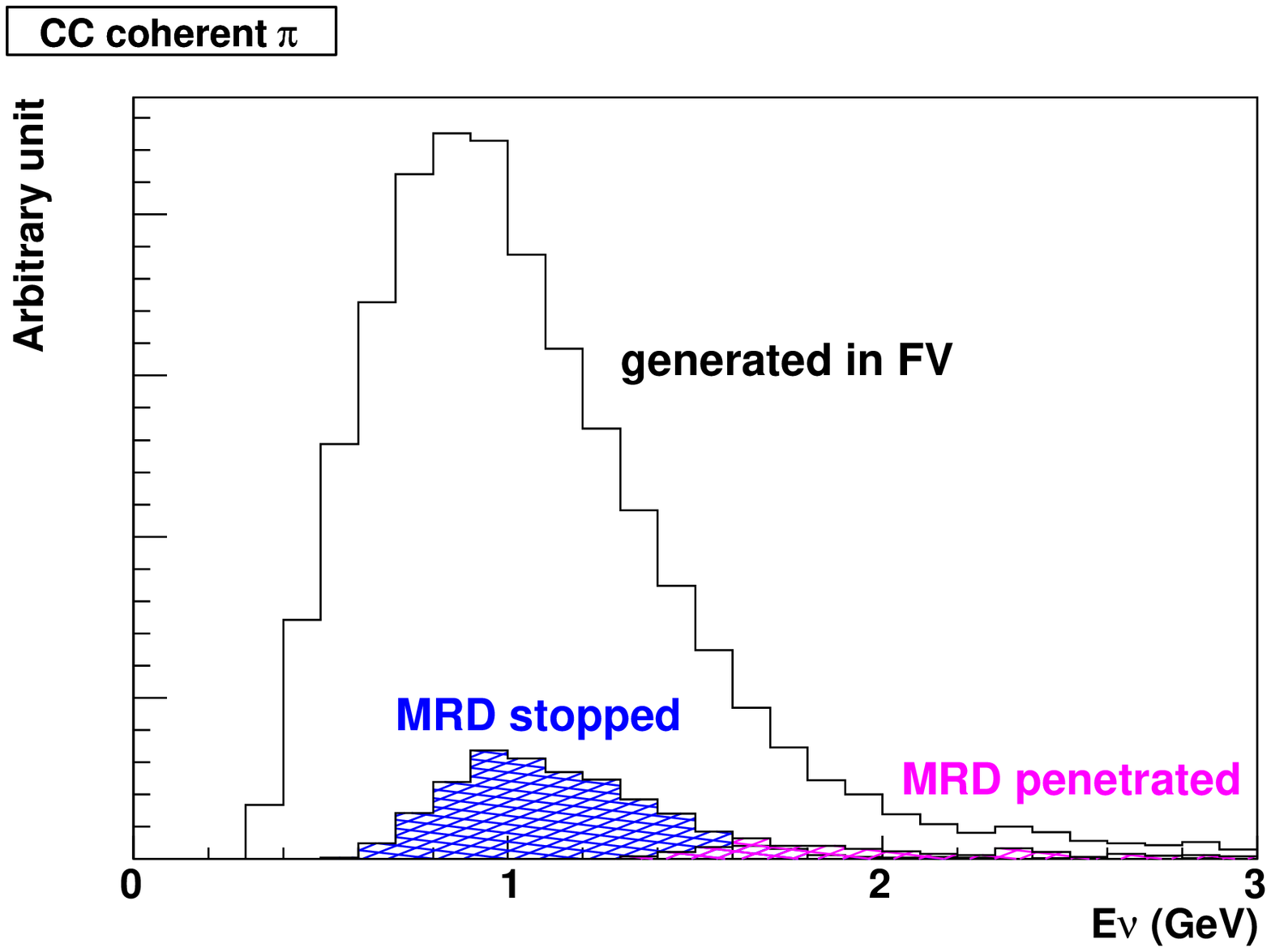}
    \includegraphics[keepaspectratio=true,height=50mm]{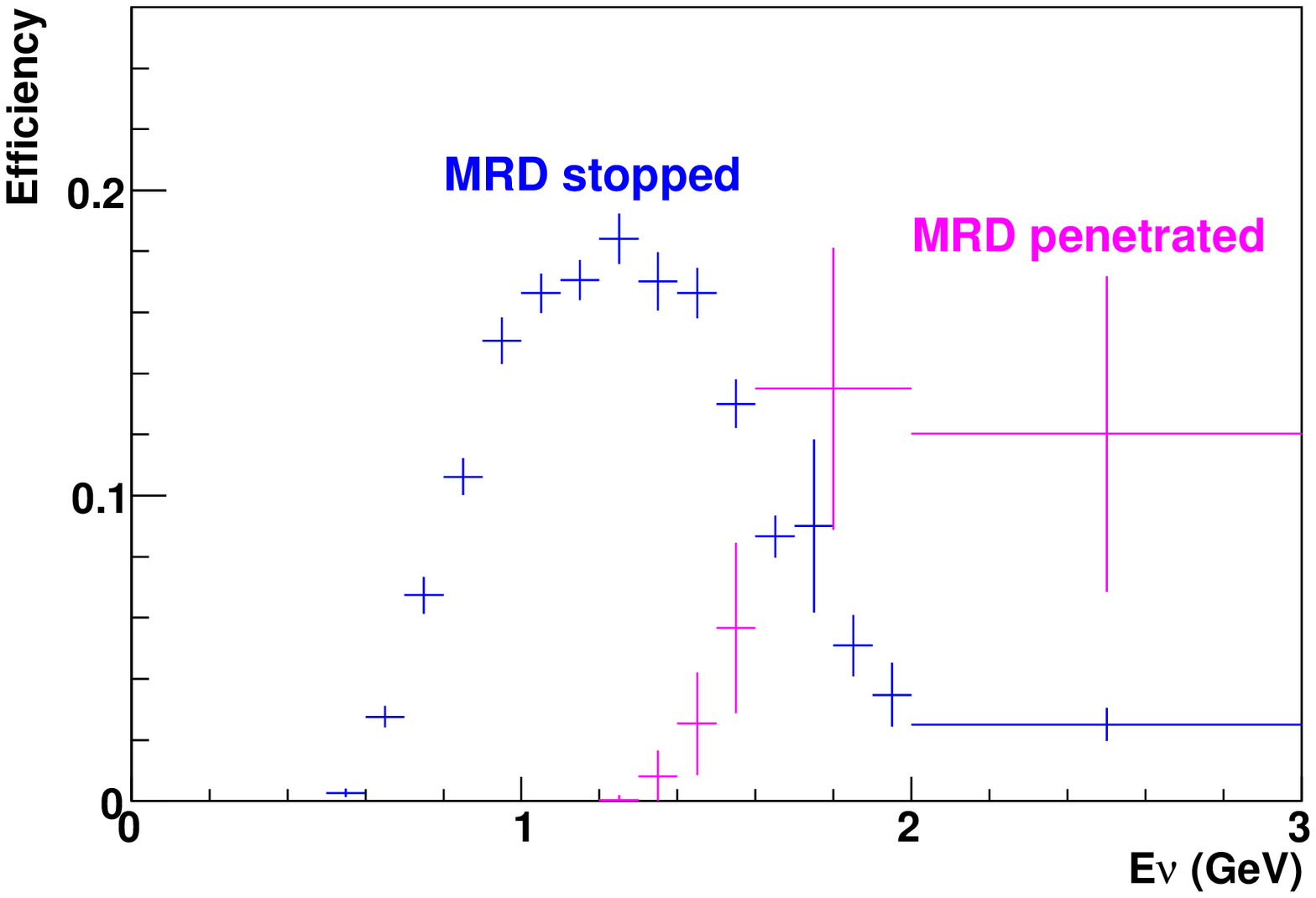}
    \caption{(Color online) Neutrino energy spectra and selection efficiencies
      as a function of neutrino energy for charged current coherent pion events.}
    \label{fig:efficiency_cccoh.eps}
    \includegraphics[keepaspectratio=true,height=50mm]{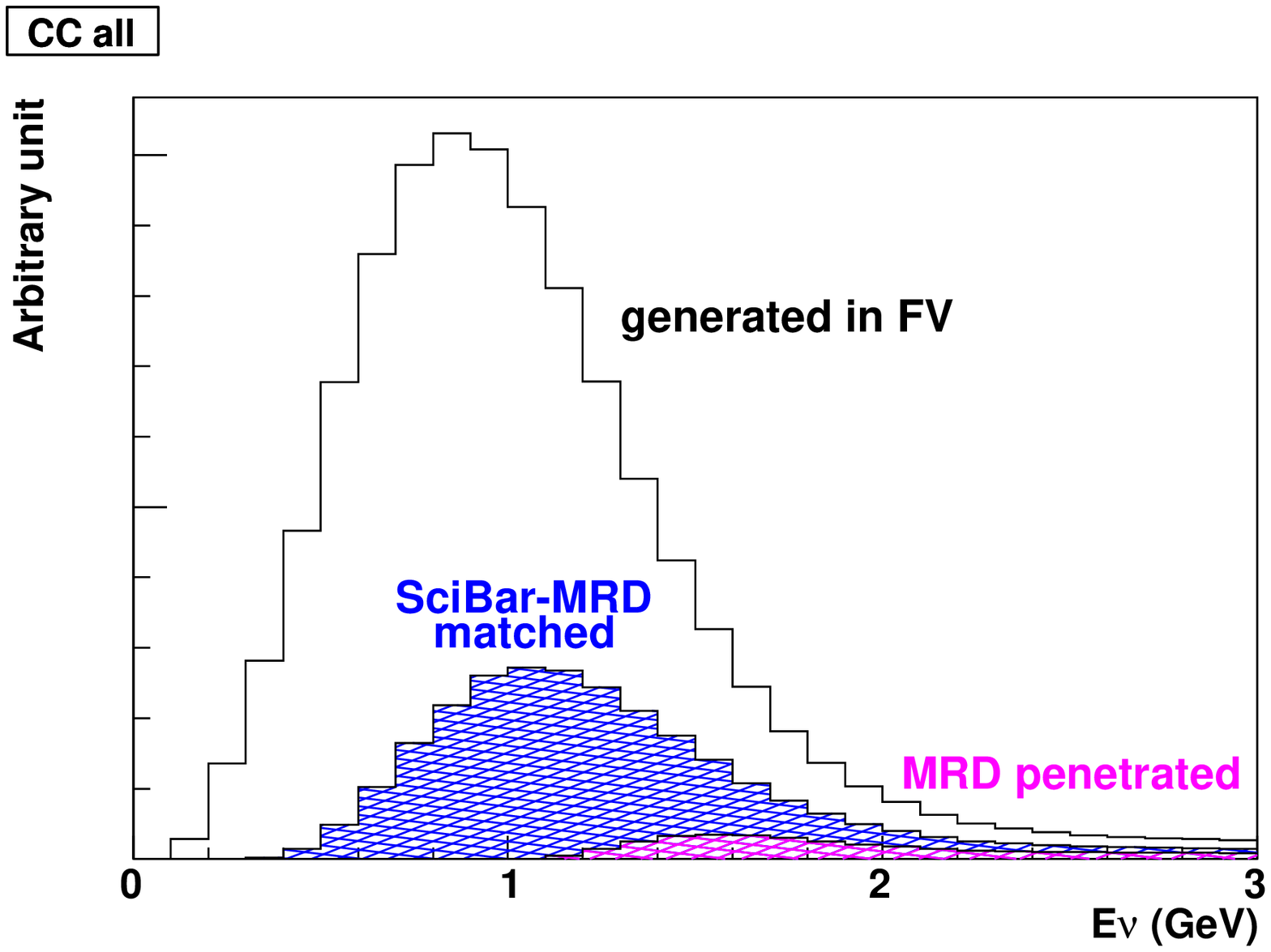}
    \includegraphics[keepaspectratio=true,height=50mm]{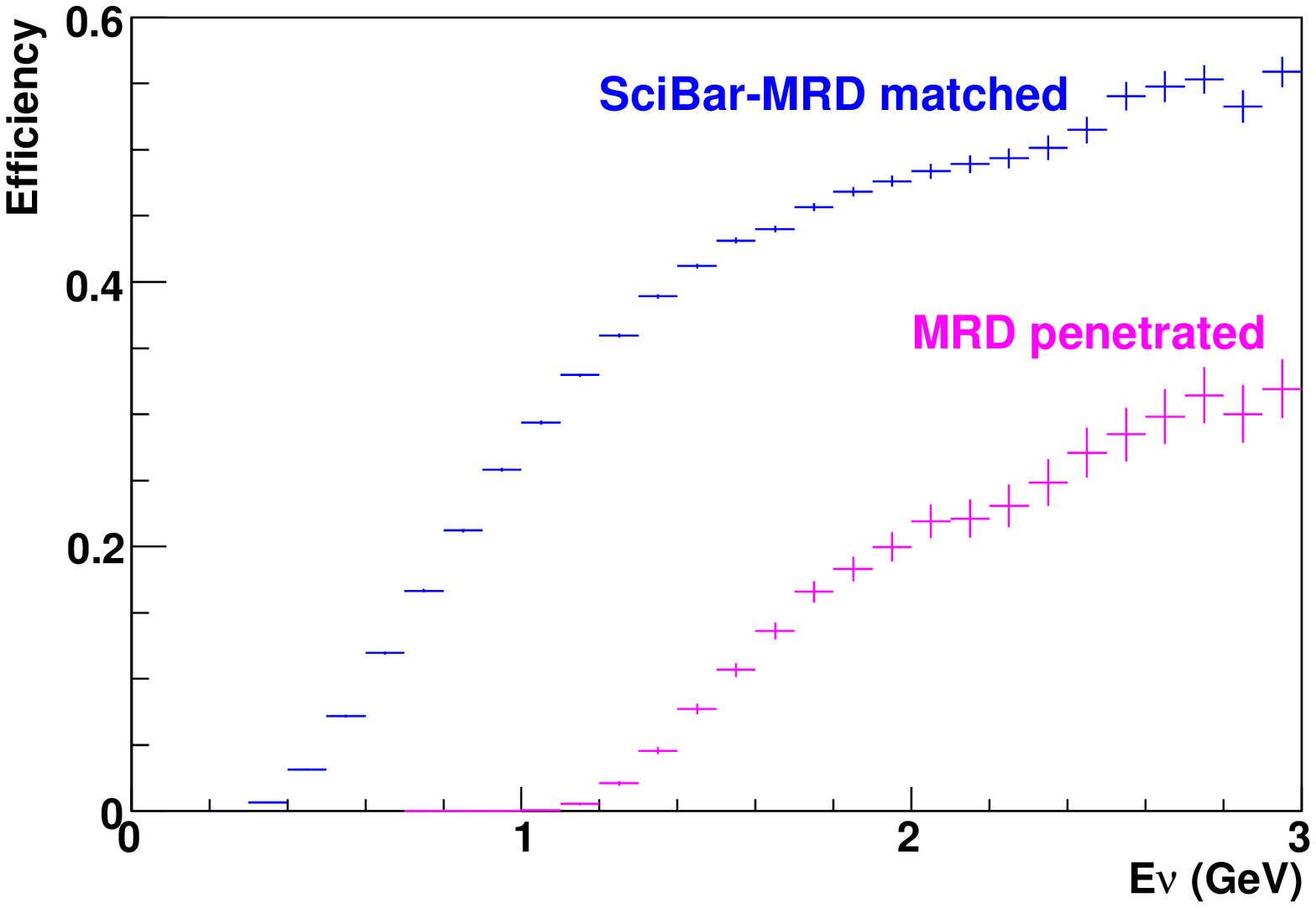}
    \caption{(Color online) Neutrino energy spectra and selection efficiencies
      as a function of neutrino energy for all charged current events.}
    \label{fig:efficiency_ccall.eps}
  \end{center}
\end{figure}

After subtracting background and correcting for the selection
efficiency, the number of charged current coherent pion candidates in the MRD-stopped
sample is measured to be $179\pm 190$(stat); this error includes the
uncertainty in the background estimation.  No evidence of charged current coherent
pion production is found in the sample.  The neutrino energy
dependence of the selection efficiency for charged current coherent pion events is
shown in Fig.~\ref{fig:efficiency_cccoh.eps}.  The mean
neutrino beam energy for true charged current coherent pion events in the sample is
estimated to be 1.1~GeV after accounting for the effects of the selection
efficiency. The RMS of the neutrino beam energy is 0.27~GeV.

The total number of charged current interactions is estimated by using the
SciBar-MRD matched sample.  We observe 30,337 SciBar-MRD matched
events. As described in section~\ref{sec:analysis}, the selection
efficiency and purity of charged current events are estimated to be 27.9\% and
92.8\%, respectively. The neutrino energy dependence of the selection
efficiency for charged current events is shown in
Fig.~\ref{fig:efficiency_ccall.eps}.  After correcting for the
efficiency and purity, the number of charged current events is measured to be
$(1.091\pm 0.006({\rm stat}))\times 10^5$.

Using this information, the ratio of the charged current coherent pion to
total charged current production cross sections is measured to be $(0.16\pm0.17({\rm stat})
^{+0.30}_{-0.27}({\rm sys}))\times 10^{-2}$ at 1.1~GeV, where the
systematic error is described later.  The result is consistent with
the non-existence of charged current coherent pion production, and hence we set an
upper limit on the cross section ratio by using the likelihood distribution
($\mathcal{L}$) which is convolved with the systematic error.
We calculate the 90\% confidence level (C.L.) upper limit (UL) using the relation
$\int_0^{\rm UL} \mathcal{L} dx / \int_0^{\infty} \mathcal{L} dx = 0.9$ to be:
\begin{eqnarray}
  \frac{\sigma(\mbox{CC coherent } \pi)}{\sigma({\rm CC})} < 0.67 \times 10^{-2}
\end{eqnarray}
at a mean neutrino energy of 1.1~GeV.

\subsubsection{MRD penetrated charged current coherent pion sample}
After subtracting background and correcting for the selection
efficiency, the number of charged current coherent pion candidates in the MRD
penetrating sample is measured to be $548\pm 254$(stat).  As in the
MRD stopping sample, this includes the uncertainty due to the
background estimation.  The mean neutrino beam energy for true charged current
coherent pion events in the sample is estimated to be 2.2~GeV after
accounting for the effects of the selection efficiency.
The RMS of the neutrino beam energy is 0.80~GeV.

Due to the higher neutrino energy in the charged current coherent pion sample, the MRD
penetrated charged current sample is chosen to estimate the number of total charged current
interactions at a similar neutrino energy. We observe 3,712 MRD
penetrated events, and the efficiency and purity of true $\nu_\mu$
charged current events are estimated to be 4.5\% and 97.5\%, respectively.
The impurity largely comes from $\overline{\nu}_\mu$ charged current events.
After correcting the efficiency and
purity, the number of charged current events is measured to be $(0.804\pm
0.013({\rm stat}))\times 10^5$. A 26\% difference between the MRD
matched and penetrated samples is found while the estimated
uncertainty due to the neutrino flux is 14\%.
However, this is expected to be a small effect on the cross section
ratio measurement.

The ratio of the charged current coherent pion to total charged current production cross
sections is measured to be $(0.68\pm 0.32({\rm stat})
^{+0.39}_{-0.25}({\rm sys}))\times 10^{-2}$ at 2.2~GeV.  The
systematic error is described later.  No significant evidence for charged current coherent
pion production is observed, and hence we set an upper limit on the cross section ratio at
90\% C.L.:
\begin{eqnarray}
  \frac{\sigma(\mbox{CC coherent } \pi)}{\sigma({\rm CC})} < 1.36 \times 10^{-2}
\end{eqnarray}
at a mean neutrino energy of 2.2~GeV.

\subsection{Systematic Uncertainties}
\label{subsec:systematic}
The sources of systematic error are divided into five categories, (i)
detector response and track reconstruction, (ii) nuclear effects,
(iii) neutrino interaction models, (iv) neutrino beam, and (v) event
selection. We vary these sources within their
uncertainties and take the resulting change in the cross section ratio
as the systematic uncertainty of the measurement.
Table~\ref{table:systematic_error} summarizes the uncertainties in the
charged current coherent pion cross section ratio for the MRD stopped and MRD
penetrated samples.  The total systematic error is
$^{+0.30}_{-0.27}\times 10^{-2}$ for the MRD stopped sample, and
$^{+0.39}_{-0.25}\times 10^{-2}$ for the MRD penetrated sample.

\begin{table}[htbp]
 \caption{Summary of the systematic errors in the charged current coherent pion cross section ratio.}
 \label{table:systematic_error}
 \begin{center}
  \begin{tabular}{lrr|rr}
    \hline \hline
    Source          & \multicolumn{2}{c|}{MRD stopped} & \multicolumn{2}{c}{MRD penetrated}  \\
                    & \multicolumn{2}{c|}{error ($\times 10^{-2}$)}
                    & \multicolumn{2}{c}{error ($\times 10^{-2}$)} \\
    \hline
    Detector response          & +0.10 & $-$0.18 & +0.18 & $-$0.18 \\
    Nuclear effect             & +0.20 & $-$0.07 & +0.19 & $-$0.09 \\
    Neutrino interaction model & +0.17 & $-$0.04 & +0.08 & $-$0.04 \\
    Neutrino beam              & +0.07 & $-$0.11 & +0.27 & $-$0.13 \\
    Event selection            & +0.07 & $-$0.14 & +0.06 & $-$0.05 \\    
    \hline
    Total                      & +0.30 & $-$0.27 & +0.39 & $-$0.25 \\
    \hline \hline
  \end{tabular}
 \end{center}
\end{table}

\subsubsection{Detector Response and Track Reconstruction}
The crosstalk of the MA-PMT was measured to be 3.15\% for adjacent
channels, with an absolute error of 0.4\%.  The single photoelectron
resolution of the MA-PMT is set to 50\% in the simulation, to
reproduce the observed $dE/dx$ distribution of cosmic muons.  The
absolute error is estimated to be $\pm$20\%.  Birk's constant of
the SciBar scintillator was measured to be $0.0208 \pm 0.0023 \ {\rm
cm/MeV}$ \cite{Hasegawa:2006am} and is varied within the
measurement error to evaluate the systematic. The hit threshold for
track reconstruction is varied by $\pm$20\%.

\subsubsection{Nuclear Effects}
We consider uncertainties in final state interactions inside the
nucleus.  This includes rescattering of nucleons and pions in the
initial target nucleus.  For pions produced by neutrino interactions,
uncertainties on the cross sections for pion absorption and pion
inelastic scattering in the nucleus are considered.  The cross section
of pion charge exchange is negligible compared with the other effects
and is hence neglected.  In the momentum range of pions from $\Delta$
decays, the cross section measurement uncertainty for both absorption
and inelastic scattering is approximately 30\% \cite{Ashery:1981tq}.
Nucleon re-scattering in the nucleus affects vertex activity. The
uncertainty in the cross section is estimated to be 10\%.  In the NEUT
simulation, the Fermi momentum of nucleons is set to 217~MeV/$c$ for
carbon.  According to electron quasi-elastic scattering data
\cite{Moniz:1971mt}, the value is approximately $221\pm 5$~MeV/$c$.
Therefore, an uncertainty of $\pm$5~MeV/$c$ is assigned.

\subsubsection{Neutrino Interaction Models}
In the NEUT simulation, we set the axial vector mass $M_A$ to 1.21
GeV/$c^2$ for both QE and resonant pion production.  The uncertainty
in this value is estimated to be approximately $\pm 0.1$~GeV/$c^2$
based on recent measurements \cite{Gran:2006jn,:2007ru}; results from
past experiments are systematically lower than the
recent measurements \cite{Bernard:2001rs}, and thus we only vary $M_A$ to
1.11 GeV/$c^2$, and take that change as the systematic error.
We consider the uncertainty in the charged current resonant $\mu^-n\pi^+ /
\mu^-p\pi^+$ cross section ratio. The uncertainty in this ratio is
estimated to be 7\% using four SciBooNE data samples described in
Section~\ref{sec:event_classification}. In addition, a disagreement of the $Q^2$ shape
is observed in the $\mu+\pi$ events with vertex activity where charged current
resonant pion production is dominant, as shown in
Fig.~\ref{fig:q2rec_after.eps}.  We estimate the systematic
uncertainty, due to evident low $Q^2$ suppression of charged current resonant pion
production even after the MC tuning, by re-weighting the true $Q^2$
distribution of charged current resonant pion events according to the observed low
$Q^2$ deficit.

\subsubsection{Neutrino Beam}
The uncertainties in secondary particle production cross sections in
proton-beryllium (p-Be) interactions, hadronic interactions in the target or horn,
and the horn magnetic field model are varied within their externally
estimated error bands. Detailed descriptions of each uncertainty
are found elsewhere \cite{AguilarArevalo:2008yp}.  Uncertainties
associated with the delivery of the primary proton beam to the
beryllium target and the primary beam optics, which result in the
overall normalization uncertainty, are not considered in this analysis
since it cancels in the cross section ratio.

\subsubsection{Event Selection}
For the event selection variable $\Delta \theta_p$, we evaluate the systematic
uncertainty in the cross section ratio by varying the cut placement.  The uncertainty
in the $\Delta \theta_p$ cut for charged current quasi-elastic rejection is estimated to be $\pm 5$ degrees.
For the other variables, we already tuned the MC simulation using the migration
parameters or considered possible systematic sources.  Therefore,
we do not include additional systematic uncertainties due to these selections.

\subsection{Discussion}
Having not observed evidence for charged current coherent pion production, we set
90\% C.L. upper limits at two different mean neutrino energies.
According to the Rein-Sehgal model~\cite{Rein:1982pf, Rein:2006di}
implemented in our simulation, the cross section ratio of charged current coherent
pion production to total charged current interactions is expected to be
2.04~$\times$10$^{-2}$.  Our limits correspond to 33\% and 67\% of the
prediction at 1.1 GeV and 2.2 GeV, respectively.  For reference, the
total charged current cross section averaged over the MRD stopped and MRD
penetrated samples are $1.05\times 10^{-38}$ cm$^2$/nucleon and $1.76\times
10^{-38}$ cm$^2$/nucleon, respectively, estimated with the MC simulation.  Our
results are consistent with the K2K result; $\sigma(\mbox{CC coherent
$\pi$})/\sigma({\rm CC})<0.60\times 10^{-2}$ at 90\% C.L. measured in
a 1.3~GeV wide-band neutrino beam.
As shown in Fig.~\ref{fig:coherent}, several recent models 
predict a considerably smaller coherent cross section,
which appears consistent with our results.

Because of the connection of the neutrino and antineutrino coherent
pion production processes in the theoretical models, it will be
interesting to repeat this analysis on SciBooNE's already collected
antineutrino data.  Most models predict a similar absolute cross
section for neutrino and antineutrino coherent pion production, which
means the ratio of charged current coherent pion events to charged current inclusive events is
expected to be larger in antineutrino data because of the reduced
total $\overline{\nu}$ charged current event rate.  Because of this, the
antineutrino search has the potential to be even more sensitive.

Furthermore, theoretical models also make concrete connections
between the charged and neutral current coherent pion production
processes.  As mentioned in Section~\ref{sec:introduction}, the
MiniBooNE Collaboration has already published an observation of neutral current
coherent pion production in the same neutrino beam as SciBooNE. The
SciBooNE neutral current coherent pion search is, therefore, also 
interesting and may shed considerable light on the behavior of this
interaction process.

%% file: conclusions.tex
\section{Conclusions}
\label{sec:conclusions}

In conclusion, we have searched for muon neutrino charged current coherent pion
production on carbon in the few GeV region using the full SciBooNE
neutrino data set of $0.99\times 10^{20}$ protons on target. No
evidence of charged current coherent pion production is found, and hence we set
90\% C.L. upper limits on the cross section ratio of charged current coherent pion
to total charged current production cross sections at $0.67\times 10^{-2}$ and
$1.36\times 10^{-2}$, at mean neutrino energies of 1.1~GeV and
2.2~GeV, respectively.

%% file: acknowledgments.tex
\section{Acknowledgments}
\label{sec:acknowledgments}

The SciBooNE Collaboration sincerely expresses our gratitude to Prof.
Yoji Totsuka who helped us to start this experiment.
We acknowledge the Physics Department at Chonnam National University,
Dongshin University, and Seoul National University for the loan of
parts used in SciBar and the help in the assembly of SciBar.
We wish to thank the Physics Departments at
the University of Rochester and Kansas State University for the loan
of Hamamatsu PMTs used in the MRD.  We gratefully acknowledge support
from Fermilab as well as various grants, contracts and fellowships
from the MEXT and JSPS (Japan), the INFN (Italy), the Ministry of Science
and Innovation and CSIC (Spain), the STFC (UK), and the DOE and NSF (USA).
This work was supported by MEXT and JSPS with the Grant-in-Aid
for Scientific Research A 19204026, Young Scientists S 20674004,
Young Scientists B 18740145, Scientific Research on Priority Areas
``New Developments of Flavor Physics'', and the global COE program
``The Next Generation of Physics, Spun from Universality and Emergence''.
The project was supported by the Japan/U.S. Cooperation Program in the field
of High Energy Physics and by JSPS and NSF under the Japan-U.S. Cooperative
Science Program. K.~H. would like to acknowledge support from JSPS.

%% file: cccoherentpi.bbl
\begin{thebibliography}{999}
\bibitem{Itow:2002rk}
  Y.~Itow,
  Nucl.\ Phys.\ Proc.\ Suppl.\  {\bf 112}, 3 (2002).

\bibitem{Harris:2004iq}
  D.~A.~Harris {\it et al.}  [MINERvA Collaboration],
  arXiv:hep-ex/0410005.

\bibitem{Adler:1964yx}
  S.~L.~Adler,
  Phys.\ Rev.\  {\bf 135}, B963 (1964).

\bibitem{Gershtein:1980vd}
  S.~S.~Gershtein, Yu.~Y.~Komachenko, and M.~Y.~Khlopov,
  Sov.\ J.\ Nucl.\ Phys.\  {\bf 32}, 861 (1980)
  [Yad.\ Fiz.\  {\bf 32}, 1663 (1980)].

\bibitem{Rein:1982pf}
  D.~Rein and L.~M.~Sehgal,
  Nucl.\ Phys.\  B {\bf 223}, 29 (1983).

\bibitem{Belkov:1986hn}
  A.~A.~Belkov and B.~Z.~Kopeliovich,
  Sov.\ J.\ Nucl.\ Phys.\  {\bf 46}, 499 (1987)
  [Yad.\ Fiz.\  {\bf 46}, 874 (1987)].

\bibitem{Paschos:2005km}
  E.~A.~Paschos, A.~Kartavtsev, and G.~J.~Gounaris,
  Phys.\ Rev.\  D {\bf 74}, 054007 (2006)
  [arXiv:hep-ph/0512139].

\bibitem{Rein:2006di}
  D.~Rein and L.~M.~Sehgal,
  Phys.\ Lett.\  B {\bf 657}, 207 (2007)
  [arXiv:hep-ph/0606185].

\bibitem{Singh:2006bm}
  S.~K.~Singh, M.~Sajjad Athar, and S.~Ahmad,
  Phys.\ Rev.\ Lett.\  {\bf 96}, 241801 (2006).

\bibitem{AlvarezRuso:2007tt}
  L.~Alvarez-Ruso, L.~S.~Geng, S.~Hirenzaki, and M.~J.~Vicente Vacas,
  Phys.\ Rev.\  C {\bf 75}, 055501 (2007)
  [arXiv:nucl-th/0701098].

\bibitem{AlvarezRuso:2007it}
  L.~Alvarez-Ruso, L.~S.~Geng, and M.~J.~Vicente Vacas,
  Phys.\ Rev.\  C {\bf 76}, 068501 (2007)
  [arXiv:0707.2172 [nucl-th]].

\bibitem{Amaro:2008hd}
  J.~E.~Amaro, E.~Hernandez, J.~Nieves and M.~Valverde,
  arXiv:0811.1421 [hep-ph].

\bibitem{Hasegawa:2005td}
  M.~Hasegawa {\it et al.}  [K2K Collaboration],
  Phys.\ Rev.\ Lett.\  {\bf 95}, 252301 (2005)
  [arXiv:hep-ex/0506008].

\bibitem{Faissner:1983ng}
  H.~Faissner {\it et al.},
  Phys.\ Lett.\  B {\bf 125}, 230 (1983).

\bibitem{Isiksal:1984vh}
  E.~Isiksal, D.~Rein, and J.~G.~Morfin,
  Phys.\ Rev.\ Lett.\  {\bf 52}, 1096 (1984).

\bibitem{AguilarArevalo:2008xs}
  A.~A.~Aguilar-Arevalo {\it et al.}  [MiniBooNE Collaboration],
  Phys.\ Lett.\  B {\bf 664}, 41 (2008)
  [arXiv:0803.3423 [hep-ex]].

\bibitem{Grabosch:1985mt}
  H.~J.~Grabosch {\it et al.}  [SKAT Collaboration],
  Z.\ Phys.\  C {\bf 31}, 203 (1986).

\bibitem{Vilain:1993sf}
  P.~Vilain {\it et al.}  [CHARM-II Collaboration],
  Phys.\ Lett.\  B {\bf 313}, 267 (1993).

\bibitem{Marage:1986cy}
  P.~Marage {\it et al.}  [BEBC WA59 COLLABORATION Collaboration],
  Z.\ Phys.\  C {\bf 31}, 191 (1986).

\bibitem{Allport:1988cq}
  P.~P.~Allport {\it et al.}  [BEBC WA59 Collaboration],
  Z.\ Phys.\  C {\bf 43}, 523 (1989).

\bibitem{Willocq:1992fv}
  S.~Willocq {\it et al.}  [E632 Collaboration],
  Phys.\ Rev.\  D {\bf 47}, 2661 (1993).

\bibitem{AguilarArevalo:2006se}
  A.~A.~Aguilar-Arevalo {\it et al.}  [SciBooNE Collaboration],
  arXiv:hep-ex/0601022.

\bibitem{Agostinelli:2002hh}
  S.~Agostinelli {\it et al.}  [GEANT4 Collaboration],
  Nucl.\ Instrum.\ Meth.\  A {\bf 506}, 250 (2003).

\bibitem{AguilarArevalo:2008yp}
  A.~A.~Aguilar-Arevalo {\it et al.}  [MiniBooNE Collaboration],
  arXiv:0806.1449 [hep-ex].

\bibitem{:2007gt}
  M.~G.~Catanesi {\it et al.},
  Eur.\ Phys.\ J.\  C {\bf 52}, 29 (2007)
  [arXiv:hep-ex/0702024].

\bibitem{:2007nb}
  I.~Chemakin {\it et al.}  [E910 Collaboration],
  Phys.\ Rev.\  C {\bf 77}, 015209 (2008)
  [Erratum-ibid.\  C {\bf 77}, 049903 (2008)]
  [arXiv:0707.2375 [nucl-ex]].


\bibitem{Hayato:2002sd}
  Y.~Hayato,
  Nucl.\ Phys.\ Proc.\ Suppl.\  {\bf 112}, 171 (2002).

\bibitem{Mitsuka:2008zz}
  G.~Mitsuka,
  AIP Conf.\ Proc.\  {\bf 981}, 262 (2008).

\bibitem{Llewellyn Smith:1971zm}
  C.~H.~Llewellyn Smith,
  Phys.\ Rept.\  {\bf 3}, 261 (1972).

\bibitem{Smith:1972xh}
  R.~A.~Smith and E.~J.~Moniz,
  Nucl.\ Phys.\  B {\bf 43}, 605 (1972)
  [Erratum-ibid.\  B {\bf 101}, 547 (1975)].

\bibitem{Gran:2006jn}
  R.~Gran {\it et al.}  [K2K Collaboration],
  Phys.\ Rev.\  D {\bf 74}, 052002 (2006)
  [arXiv:hep-ex/0603034].

\bibitem{:2007ru}
  A.~A.~Aguilar-Arevalo {\it et al.}  [MiniBooNE Collaboration],
  Phys.\ Rev.\ Lett.\  {\bf 100}, 032301 (2008)
  [arXiv:0706.0926 [hep-ex]].

\bibitem{Rein:1980wg}
  D.~Rein and L.~M.~Sehgal,
  Annals Phys.\  {\bf 133}, 79 (1981).

\bibitem{Berger:2007rq}
  C.~Berger and L.~M.~Sehgal,
  Phys.\ Rev.\  D {\bf 76}, 113004 (2007)
  [arXiv:0709.4378 [hep-ph]].

\bibitem{Kuzmin:2003ji}
  K.~S.~Kuzmin, V.~V.~Lyubushkin, and V.~A.~Naumov,
  Mod.\ Phys.\ Lett.\  A {\bf 19}, 2815 (2004)
  [Phys.\ Part.\ Nucl.\  {\bf 35}, S133 (2004)]
  [arXiv:hep-ph/0312107].

\bibitem{Rein:1987cb}
  D.~Rein,
  Z.\ Phys.\  C {\bf 35}, 43 (1987).

\bibitem{Kitagaki:1986ct}
  T.~Kitagaki {\it et al.},
  Phys.\ Rev.\  D {\bf 34}, 2554 (1986).

\bibitem{Singh:1998ha}
  S.~K.~Singh, M.~J.~Vicente-Vacas, and E.~Oset,
  Phys.\ Lett.\  B {\bf 416}, 23 (1998)
  [Erratum-ibid.\  B {\bf 423}, 428 (1998)].

\bibitem{Gluck:1998xa}
  M.~Gluck, E.~Reya, and A.~Vogt,
  Eur.\ Phys.\ J.\  C {\bf 5}, 461 (1998)
  [arXiv:hep-ph/9806404].

\bibitem{Bodek:2003wd}
  A.~Bodek and U.~K.~Yang,
  arXiv:hep-ex/0308007.

\bibitem{Nakahata:1986zp}
  M.~Nakahata {\it et al.}  [KAMIOKANDE Collaboration],
  J.\ Phys.\ Soc.\ Jap.\  {\bf 55}, 3786 (1986).

\bibitem{Sjostrand:1993yb}
  T.~Sjostrand,
  Comput.\ Phys.\ Commun.\  {\bf 82}, 74 (1994).

\bibitem{Salcedo:1987md}
  L.~L.~Salcedo, E.~Oset, M.~J.~Vicente-Vacas, and C.~Garcia-Recio,
  Nucl.\ Phys.\  A {\bf 484}, 557 (1988).

\bibitem{Ashery:1981tq}
  D.~Ashery, I.~Navon, G.~Azuelos, H.~K.~Walter, H.~J.~Pfeiffer, and F.~W.~Schleputz,
  Phys.\ Rev.\  C {\bf 23}, 2173 (1981).

\bibitem{Rowe:1978fb}
  G.~Rowe, M.~Salomon, and R.~H.~Landau,
  Phys.\ Rev.\  C {\bf 18}, 584 (1978).

\bibitem{Nitta:2004nt}
  K.~Nitta {\it et al.},
  Nucl.\ Instrum.\ Meth.\  A {\bf 535} (2004) 147
  [arXiv:hep-ex/0406023].

\bibitem{Yoshida:2004mh}
  M.~Yoshida {\it et al.},
  IEEE Trans.\ Nucl.\ Sci.\  {\bf 51} (2004) 3043.

\bibitem{Buontempo:1994yp}
  S.~Buontempo {\it et al.},
  Nucl.\ Instrum.\ Meth.\  A {\bf 349} (1994) 70.

\bibitem{Heikkinen:2003sc}
  A.~Heikkinen, N.~Stepanov, and J.~P.~Wellisch,
{\it In the Proceedings of 2003 Conference for Computing in High-Energy and Nuclear Physics (CHEP 03), La Jolla, California, 24-28 Mar 2003, pp
MOMT008}
  [arXiv:nucl-th/0306008].

\bibitem{Birks:1964aa}
  J.~Birks,
  {\it Theory and Practice of Scintillation Counting},
  Pergamon Press, 1964.

\bibitem{Hasegawa:2006am}
  M.~Hasegawa,
  Ph.D.~thesis, Kyoto University, 2006.


\bibitem{Maesaka:2005aj}
  H.~Maesaka,
  Ph.D.~thesis, Kyoto University, 2005.

\bibitem{Baker:1983tu}
  S.~Baker and R.~D.~Cousins,
  Nucl.\ Instrum.\ Meth.\  {\bf 221}, 437 (1984).


\bibitem{Moniz:1971mt}
  E.~J.~Moniz, I.~Sick, R.~R.~Whitney, J.~R.~Ficenec, R.~D.~Kephart, and W.~P.~Trower,
  Phys.\ Rev.\ Lett.\  {\bf 26}, 445 (1971).

\bibitem{Bernard:2001rs}
  V.~Bernard, L.~Elouadrhiri, and U.~G.~Meissner,
  J.\ Phys.\ G {\bf 28}, R1 (2002)
  [arXiv:hep-ph/0107088].


\end{thebibliography}
